\documentclass[twocolumn]{aastex62}
\pdfoutput=1 
\usepackage{amsmath,amstext,amssymb,mathtools,graphicx,color}
\usepackage[T1]{fontenc}
\usepackage{apjfonts}
\usepackage[figure,figure*]{hypcap}
\usepackage{chngpage}

\usepackage{graphicx}	
\usepackage{amsmath}	
\usepackage{amssymb}	
\usepackage{xcolor}

\definecolor{PineGreen}{HTML}{019286}

\shorttitle{Shocks in Shock Breakout}
\shortauthors{Fryer et al.}
\graphicspath{{./}{figures/}}

\begin{document}

\title{The Role of Inhomogeneities in Supernova Shock Breakout Emission}

\correspondingauthor{Chris L. Fryer}
\email{fryer@lanl.gov}

\author[0000-0003-2624-0056]{Chris L. Fryer}
\affiliation{Center for Theoretical Astrophysics, Los Alamos National Laboratory, Los Alamos, NM, 87545, USA}
\affiliation{Computer, Computational, and Statistical Sciences Division, Los Alamos National Laboratory, Los Alamos, NM, 87545, USA}
\affiliation{The University of Arizona, Tucson, AZ 85721, USA}
\affiliation{Department of Physics and Astronomy, The University of New Mexico, Albuquerque, NM 87131, USA}
\affiliation{The George Washington University, Washington, DC 20052, USA}

\author[0000-0003-1087-2964]{Christopher~J. Fontes}
\affiliation{Center for Theoretical Astrophysics, Los Alamos National Laboratory, Los Alamos, NM, 87545, USA}
\affiliation{X Computational Physics Division, Los Alamos National Laboratory, Los Alamos, NM, 87545, USA}

\author[0000-0003-1323-866X]{James S. Warsa}
\affiliation{Computer, Computational, and Statistical Sciences Division, Los Alamos National Laboratory, Los Alamos, NM, 87545, USA}

\author[0000-0002-5499-953X]{Pete W. A. Roming}
\affiliation{Southwest Research Institute Space Science and Engineering Division, 6220 Culebra Road San Antonio, TX 78238}

\author[0000-0002-7083-3038]{Shane X. Coffing}
\affiliation{Center for Theoretical Astrophysics, Los Alamos National Laboratory, Los Alamos, NM, 87545, USA}
\affiliation{Center for Laser Experimental Astrophysical Research, University of Michigan, Ann Arbor, MI, 48708, USA}

\author[0000-0002-7208-7681]{Suzannah R. Wood}
\affiliation{Center for Theoretical Astrophysics, Los Alamos National Laboratory, Los Alamos, NM, 87545, USA}
\affiliation{Computer, Computational, and Statistical Sciences Division, Los Alamos National Laboratory, Los Alamos, NM, 87545, USA}
\affiliation{X Theory Division, Los Alamos National Laboratory, Los Alamos, NM, 87545, USA}

\begin{abstract}

The breakout of a supernova blast wave from its progenitor star provides strong constraints on the star and its immediate surroundings.  These surroundings are shaped by mass loss from the star and can include a wide variety of inhomogeneities.  Here we present results of multi-dimensional radiation-hydrodynamics calculations of the interactions of the supernova blast wave with inhomogeneities in the immediate surroundings of a massive Wolf-Rayet star, calculating the effect these interactions have on the shock breakout signal from supernovae.  

\end{abstract}



\section{Introduction} \label{sec:intro}

In core-collapse supernovae, the blastwave launched from the collapsed core of a massive star powers through and breaks out of the collapsing star.  Until this blast breaks out of the star, the radiation in the shock is trapped in the flow.  After breakout, the radiation is able to leak out of the blastwave, producing an early burst of high-energy (ultraviolet and X-ray) photons.  Initially proposed as a mechanism to produce gamma-ray bursts~\citep{1968CaJPh..46..476C}, the properties of these radiation-driven shocks have been probed extensively in both analytic and simulation studies~\citep[for a review, see][]{2017hsn..book..967W}.  

Despite a continuous stream of theoretical studies, undisputed observations of this phenomenon have remained elusive.  To date, one of the best potential observations of shock breakout has been the serendipitous SWIFT observation of SN 2008D~\citep{2008Natur.453..469S}.  The transient community has developed a number of wide-field survey telescopes that are discovering supernovae at an unprecedented rate and these telescopes will allow both a higher rate of discovery but the ability to follow the cooling of the shock breakout~\citep[again, see the review by][]{2017hsn..book..967W}.  Proposed wide-field ultraviolet detectors [e.g. http://space.gov.il/en/node/1129, \cite{2014AJ....147...79S} or the Astrophysical Transient Observatory, \cite{2012ApJ...751...92R}] could dramatically increase what we will learn from shock breakout observations.  

Simple analytic estimates of shock breakout provide direct relations between the energy/duration of the shock breakout and the stellar radius.  Simulations of shock breakout have shown that the picture is much more complex than these analytic estimates predict.  For example, many of the analytic studies focus on the energy loss from shock breakout.  However, during shock breakout, momentum deposition from the radiation can also alter the flows, changing the breakout signal~\citep[for a review, see][]{2011AstL...37..194B,2013MNRAS.429.3181T}.  To capture this physics, astronomers must include the interaction of radiation with matter.

The complex opacity structure also alters the nature of shock breakout.  \cite{2013ApJS..204...16F} found that the duration of the shock breakout signal was dramatically broadened because the opacity (and hence photosphere) varies with photon energy and different wavelength photons break out of the shock at different times.  These calculations assumed the emission was in local thermodynamic equilibrium.  But, in many cases, this is not the case.  Non local thermodynamic equilibrium conditions can also alter the emission~\citep{1978ApJ...223L.109K,2007ApJ...664.1026W,2010ApJ...719..881S}.

Another suite of studies studied the dependence of the shock breakout emission on the nature of the star and its immediate surroundings.  \cite{2015ApJ...805...98B} found that the mass (and density) of the stellar envelope can alter the shock breakout signal.  \cite{2017ApJ...845..103L} argued that the structure of the transition profile connecting the star to the stellar wind can also alter the signal.  Interactions with the circumstellar medium can alter the shock breakout emission significantly~\citep[e.g.][]{2017ApJ...850..133D}.  

All of this work assumed a smooth density profile connecting the the massive star to its circumstellar medium.  However, there is strong evidence that stellar mass-loss is very different than the constant or slowly varying mass loss rates assumed in producing the simple $r^{-2}$ density profiles in many of the shock breakout studies.  Explosive shell burning, opacity-driven instabilities and pressure waves can all produce bursts of mass ejection in a star, producing inhomogeneities in the circumstellar medium including both shells or clumpy media~\citep{2006ApJ...647.1269F,2014ApJ...792L...3H,2016MNRAS.458.1214Q}.  Studies of line-driven winds also indicate that even these, relatively quiescent outflows, can produce large inhomogeneities in the circumstellar medium~\citep{1984ApJ...284..337O, 2008A&ARv..16..209P, 2018Natur.561..498J,2019MNRAS.485..988O}.  In this paper, we study the effect these inhomogeneities on the shock breakout emission.  Section~\ref{sec:codes} describes our simulation methods and Section~\ref{sec:shockphys} describes the basic physics behind our calculations.  The blastwave evolution as it shocks against these inhomogeneities is described in Section~\ref{sec:radhydro} and the spectra and light-curves of the breakout emission is described in Section~\ref{sec:spectra}.

\section{Simulation Tools}
\label{sec:codes}

In these calculations, we use a higher order radiation-transport code to model the radiation flow coupled with shock interactions.  This calculation uses a coarse, 24-energy group resolution for the photons to model energy transport.  We post-process these calculations with a 1500-energy group ray-trace code.  Both calculations use opacities from the LANL dataset.  We describe these methods in further detail below.

\subsection{Radiation Hydrodynamics Simulations}

The hydrodynamics method used in our simulations leverages the adaptive mesh refinement framework developed for the RAGE code~\citep{2008CS&D....1a5005G}.  The hydrodynamics package is a cell-based adaptive mesh refinement scheme using a two-shock approximate Riemann solver.  The code has been verified against a variety of analytic test problems, the most relevant for this problem being the Sedov blast wave~\citep{2008CS&D....1a5005G}.  It has been used extensively in the laboratory experimental community and its results have been compared to a number of other codes in this community~\citep[e.g.][]{2013HEDP....9...63F,2014PhRvE..90c3107F}.  Many of these comparisons included validation tests against laboratory experiments and we will discuss the specific tests of radiation-hydrodynamics with our current transport scheme in Section~\ref{sec:shockphys}.  It has also been compared to a wide variety of astrophysics codes~\citep[e.g.][]{2014JCoPh.275..154J}.  This hydrodynamics scheme, coupled with flux-limited diffusion, has modeled a number of astrophysical transients~\citep{2009ApJ...707..193F,2010AIPC.1294...70F,2010ApJ...725..296F,2013ApJ...762L...6W,2013ApJS..204...16F,2013ApJ...768...95W,2013ApJ...773L...7F,2013ApJ...774...64W,2013ApJ...777...99W,2013ApJ...777..110W,2013ApJ...778...17W,2014ApJ...781..106W,2014ApJ...797....9W,2014ApJ...797...97S,2015ApJ...805...44S}.  

In this paper, we instead use a higher-order, multigroup, S$_N$ (discrete ordinates) radiation transport method~\citep{1950ratr.book.....C}. The angular dependence of the radiation intensity is discretized via a collocation method for a set of directions that correspond to the nodes and weights of a quadrature set that integrates a function over the surface of a sphere.
The multigroup S$_N$ radiation equations are solved separately from the hydrodynamics equations in
a time step. The radiation equations are coupled to expressions for the energy balance in the material, taking into account the energy of both the electron and the ions. The hydrodynamic equations are coupled to the radiation field through the energy and momentum deposited by the field into the material.
Because the radiation-hydrodynamics equations are being calculated in the Eulerian reference frame, care must be taken to account for the motion of the fluid in the radiation equations, including relativistic effects and to ensure the equilibrium and diffusion limits \citep{granlibakken2004}.
Local thermodynamic equilibrium is usually assumed in these simulations such that the emissivity and absorptivity is Planckian.
 
\newcommand{\bA}[1]{\ensuremath{\mathbf{#1}}}
\newcommand{\bF}[1]{\ensuremath{\bm{\widehat{#1}}}}
\newcommand{\bO}[1]{\ensuremath{\mathbf{\widetilde{#1}}}}
\newcommand{\nv}{\ensuremath{\hat{n}}}
\newcommand{\gr}{\ensuremath{\bA{\nabla}}}
\newcommand{\on}{\ensuremath{\hat{\Omega}}}
\newcommand{\rv}{\ensuremath{r}}
\newcommand{\rp}{\ensuremath{r}^{\prime}}
\newcommand{\REQ}[1]{(\ref{eq:#1})}
\newcommand{\LEQ}[1]{\label{eq:#1}}

The key unknowns in the radiation equations are the radiation intensity $I(\rv,\nu)$, 
and the temperatures, $T_e=T_e(\rv)$ and $T_i=T_i(\rv)$.
The total, scattering, and absorption opacities, $\kappa_t$ $\kappa_s$ and $\kappa_a$, 
respectively, specific heats $C_{ve}$ and $C_{vi}$, sources $Q_e$, $Q_i$, $Q_i$, electron-ion coupling coefficient $\alpha$, and density $\rho$ are implicitly functions of time, $t$.
\begin{subequations}
\LEQ{rad}
\begin{equation}
\begin{aligned}
  \frac{1}{c} \frac{\partial I}{\partial t} 
+ \on \cdot \nabla I 
&+  \kappa_{t}(T_e,\nu) I(\rv,\on,\nu)
= \frac{\kappa_{s}(T_e,\nu)}{4 \pi} \phi(\rv,\nu) \\
&+ \kappa_{a}(T_e,\nu) B(T_e,\nu)
+ Q(\rv,\on,\nu),
\end{aligned}
\end{equation}
\begin{equation}
\begin{aligned}
  \rho C_{ve}(T_e) \frac{\partial T_e}{\partial t} 
&= \int_{0}^{\infty} d\nu \kappa_{a}(T_e,\nu) \left[ \phi(\rv,\nu) - 4\pi B(T_e,\nu) \right] \\
&+ \alpha \left(T_i - T_e\right) + Q_e(\rv,\nu),
\end{aligned}
\end{equation}
\begin{equation}
  \rho C_{vi}(T_i) \frac{\partial T_i}{\partial t} 
= \alpha \left(T_e - T_i\right) + Q_i(\rv,\nu),
\end{equation}
where
\begin{equation}
  \phi(\rv,\nu) 
= \int_{4\pi} \, d\Omega \, I(\rv,\on,\nu), 
\end{equation}
\begin{equation}
  B(T_e,\nu) = \frac{2 h \nu^3}{c^2} \left(e^{h\nu/k T_e} - 1\right)^{-1},
\end{equation}
and
\begin{equation}
\int_{0}^{\infty} d\nu B(T_e,\nu) = a c T_e^{4}.
\end{equation}
\end{subequations}
The dependence on the spatial location $\rv$ will now be suppressed.
Introduce a time discretization index $n$ and use backward--Euler differencing
for the time derivatives with time step $\Delta t_n$, 
the multigroup approximation in frequency with index $g$, and
angular quadrature index $m$.
Let $C_{ve}^{n} = C_{ve}(T_e^{n})$, $C_{vi}^{n} = C_{vi}(T_i^{n})$, 
$\kappa_{t,g}^{n} = \kappa_{t,g}(T_e^{n})$, 
$\kappa_{s,g}^{n} = \kappa_{s,g}(T_e^{n})$,
$\kappa_{a,g}^{n} = \kappa_{a,g}(T_e^{n})$, 
and $\rho^n$ be the density at time step $n$.
in \REQ{rad}
\begin{subequations}
\LEQ{rad-diff} 
\begin{equation}
\LEQ{rad-trans}
\begin{aligned}
  \frac{1}{c} \frac{I_{m,g}^{n+1} -I_{m,g}^{n}}{\Delta t_n} 
+ \on \cdot \nabla I_{m,g}^{n+1} 
&+  \kappa_{t,g}^{n} I_{m,g}^{n+1}
= \frac{\kappa_{s,g}^{n}}{4 \pi} \phi_{g}^{n+1} \\
&+ \kappa_{a,g}^{n} B_g(T_e^{n+1})
+ Q_{m,g}^{n+1},
\end{aligned}
\end{equation}
\begin{equation}
\LEQ{electron-energy}
\begin{aligned}
\rho^n C_{ve}^{n} \frac{\delta T_e}{\Delta t_n} 
&= \sum_{g} \kappa_{a,g}^{n} \left[ \phi_g^{n+1} - 4\pi B_g(T_e^{n+1}) \right] \\
&+ \alpha \left(T_i^{n+1} - T_e^{n+1}\right) + Q_e^{n+1},
\end{aligned}
\end{equation}
\begin{equation}
\LEQ{ion-energy}
  \rho^n C_{vi}^{n} \frac{\delta T_i}{\Delta t} 
= \alpha \left(T_e^{n+1} - T_i^{n+1}\right) + Q_i^{n+1},
\end{equation}
where
\begin{equation}
\phi_{g}^{n+1} = \sum_{m} w_{m} I_{m,g}^{n+1}, 
  \quad \sum_{m} w_{m} = 4 \pi,
\end{equation}
and
\begin{equation}
B_{g}(T_e) = \int_{\nu_{g-1}}^{\nu_{g}} d\nu B(T_e,\nu).
\end{equation}
\end{subequations}
Letting $T_e^{n+1} = T_e^{n} + \delta T_e$, linearize
\begin{equation}
\LEQ{planck-linearization}
B_g^{n+1} = B_g^{n} + b_g^{n} \delta T_e,
\end{equation}
where 
\[
B_g^{n+1} = B_g(T_e^{n+1}), \quad B_g^{n} = B_g(T_e^{n}), \quad  \text{and} \quad
b_g^{n} = \left. \dfrac{\partial B_g}{\partial T_e} \right\vert_{\mbox{$T_e^{n}$}}.
\]
Solve \REQ{rad-diff} by reducing the system as follows.
First, use \REQ{ion-energy} to find
\begin{subequations}
\begin{equation}
\LEQ{Ti}
T_i^{n+1} = \gamma T_i^{n} + \left( 1-\gamma\right ) T_e^{n+1} +
\left( \dfrac{1-\gamma}{\alpha}\right)Q_i^{n+1},
\end{equation}
where
\begin{equation}
\gamma = \dfrac{\rho^n C_{vi}^{n}}{\rho^n C_{vi}^{n} + \alpha \Delta t_n}.
\end{equation}
\end{subequations}
Substitute this and \REQ{planck-linearization} into \REQ{electron-energy}
\begin{equation}
\LEQ{delta-Te}
\begin{aligned}
&\left[ 
\dfrac{\rho^n C_{ve}^{n}}{\Delta t_n} + 4 \pi \sum_{k} \kappa_{a,k}^{n} b_k^{n} + \alpha \gamma 
\right] \Delta t_{n_e} 
= 
\sum_{k} \kappa_{a,k}^{n} \left( \phi_k^{n+1} - 4\pi B_k^{n} \right) \\
&+ \alpha \gamma \left( T_i^{n} - T_e^{n} \right) +
\left(1-\gamma\right)Q_i^{n+1} + Q_e^{n+1}.
\end{aligned}
\end{equation}
Solve for $\delta T_e$ and substitute into \REQ{planck-linearization},
then substitute the result of that in the right hand side of the 
radiation transport equation \REQ{rad-trans},
\begin{subequations}
\LEQ{rad-trans-final}
\begin{equation}
\LEQ{trans-final}
  \on \cdot \nabla I_{m,g}^{n+1} 
+  \left( \kappa_{t,g}^{n} + \tau \right) I_{m,g}^{n+1}
= \dfrac{\kappa_{s,g}^{n}}{4 \pi} \phi_{g}^{n+1}
+ \dfrac{\chi_g}{4 \pi} \sum_{k} \nu \kappa_{a,k}^{n} \phi_{k}^{n+1} + q_{g},
\end{equation}
where $\tau = \dfrac{1}{c \Delta t_n}$,
\begin{equation}
\nu = 
\dfrac{ {\displaystyle 4 \pi \sum_{k} \kappa_{a,k}^{n} b_{k}^{n}}}
      {\dfrac{\rho^n C_{ve}^{n}}{\Delta t_n} 
      + {\displaystyle 4 \pi \sum_{k} \kappa_{a,k}^{n} b_{k}^{n}} 
      + \alpha \gamma},
\end{equation}
\begin{equation}
\chi_g = 
\dfrac{\kappa_{a,g}^{n} b_{g}^{n}}
      {{\displaystyle \sum_{k} \kappa_{a,k}^{n} b_{k}^{n}}},
\end{equation}
and
\begin{equation}
\begin{aligned}
q_g 
& = \dfrac{\nu \chi_g}{4 \pi} 
  \left[ \alpha \gamma \left(T_i^{n} - T_e^{n}\right) 
      - {\displaystyle 4 \pi \sum_{k} \kappa_{a,k}^{n} B_{k}^{n}}
      + \left(1-\gamma\right)Q_i^{n+1} + Q_e^{n+1}
  \right] \\
& + \kappa_{a,g}^{n} B_{g}^{n} 
  + Q_{m,g}^{n+1}
  + \tau I_{m,g}^{n}
\end{aligned}
\end{equation}
\end{subequations}

The expressions in \REQ{trans-final} are discretized in space with a linear-discontinuous, or dG(1), finite element method. 
An important consideration is that the discrete equations preserve the asymptotic thick-diffusion limit. The dG(1) method used for the simulations described in this paper preserve that limit; for a 1-dimensional example see \citep{1996JCoPh.128..445M}.
The discretized equations are solved for the $I_{m,g}^{n+1}$ at every spatial degree of freedom
in the problem, then used to compute $\phi_{g}^{n+1}$, and stored for the next time step.
A Krylov iterative method preconditioned with linear multi-frequency gray acceleration~\citep{TILL2018931} is used to calculate the solution. 
This in turn is substituted into \REQ{delta-Te} to find $\Delta t_{n_e}$ 
and then $T_e^{n+1} = T_e^{n} + \delta T_e$. 
The last operation is to use the newly computed value for $T_e^{n+1}$ in \REQ{Ti} 
to find $T_i^{n+1}$.
Note that all of these expressions are computed for every degree-of-freedom 
associated the spatial discretization. 
The (cell-average) radiation energy density and momentum~\citep[correct to $O(v/c)$,][]{1999ApJ...521..432L} are calculated from $I_{m,g}^{n+1}$ and used in the (cell-centered) hydrodynamics equations for the next time step.
A second-order projection/interpolation scheme is combined with the RAGE adaptive mesh refinement capability to spatially adapt the discontinuous radiation field and temperatures.

This S$_N$ scheme has been compared in a radiation-hydrodynamics setting to an implicit Monte Carlo transport scheme coupled to our same hydrodynamics package.  In general, there is good agreement between these two methods~\citep{2020HEDP...3500738F}, with only mild differences when modeling energy transport across material boundaries.  These differences are minimal and believed to be due to artificial transport caused by the simplified re-emission term in the implicit Monte Carlo package.

It is worth comparing this transport scheme to others used in the supernova light-curve and stellar wind communities.  A number of codes exist that run typically in pure transport mode (assuming that the material is homologously expanding).  These schemes include Monte-Carlo solutions~\citep{2006ApJ...651..366K,2013ApJS..209...36W}, discrete ordinate~\citep{1991JQSRT..46...81H}, and a variety of moment closure techniques (integrating over angle) like flux-limited diffusion or variable Eddington factor methods~\citep{1993ApJ...412..731E}.  To date, the bulk of the radiation-hydrodynamics calculations have used these moment closure techniques~\citep{2013MNRAS.429.3181T, 2013ApJS..204...16F}.  These full transport schemes are necessary in conditions where the optical depth is greater than one.  In some wind calculations, the conditions are such that a ray-by-ray approach with a flux-weighted opacity sufficient to capture the physics~\citep{Sander17,Sander18,Sander20}.  Our conditions are sufficiently optically thick that a full transport scheme is more appropriate, but we use a ray trace to calculate the detailed spectra using the temperatures set by our full transport method (see Section~\ref{sec:raytrace}).

The radiation couples with the hydrodynamics both in the energy and momentum equations through the photon absorption and emission terms.  For these calculations, we use an ideal gas equation of state. At the low densities of the stellar wind material, electrons dominate the material pressure and this ideal-gas approximation is very accurate.  In addition, for these calculations, radiation pressure often dominates the total pressure and small errors in the exact pressure are negligible compared to this total pressure.

\subsection{Post-process for Spectra}
\label{sec:raytrace}

Our radiation-hydrodynamics calculations resolve the photon energy with 24 groups (we performed a convergence study of groups to confirm this coarse group structure captured the energy and momentum deposition).  Although this coarse group structure is sufficient to calculate the energy transport, it does not produce detailed spectra.  In our post-process, we are able to model a much more resolved spectrum.  The post-process assumes the temperatures and densities from the radiation-hydrodynamics calculations are accurate and uses a ray trace to calculate the emission and absorption from ejecta.  The contribution from each zone in the calculation to the emission ($L_i$) is set by:
\begin{equation}
    L_i(\nu) = r \pi r_i^2 \kappa(\nu,\rho^n_i,T_i) B(\nu,T_i) e^{-\tau_i}
\end{equation}
where $r_i$ is the radius of the zone  $i$, $\kappa(\nu,\rho_i,T_i)$ is the opacity at frequency $\nu$ for a zone density $\rho_i$ and temperature $T_i$, $B(\nu,T_i)$, is the blackbody emission and $\tau_i$ is calculated by numerically integrating inward from the observer:
\begin{equation}
    \tau_i=\sum_{r_{\rm outer}}^i \kappa(\nu,\rho_i,T_i) \rho_i dr_i
\end{equation}
where $dr_i$ is the zone size.  We use the full opacities described in Section~\ref{sec:opacities}.  From this, we can calculate detailed spectra and light-curves.

\subsection{Opacities and Opacity Implementation}
\label{sec:opacities}

For the present calculations, we use multigroup opacities generated with the TOPS code\footnote{http://aphysics2.lanl.gov/opacity/lanl} from monochromatic data contained in the tabular OPLIB database~\citep{colgan2016,hakel2004,hakel2006,kilcrease2015}.  This database was produced with the Los Alamos suite of atomic physics codes \cite[see][for an overview]{fontes2017}, which has been used to calculate spectral quantities for a variety of astrophysical applications, e.g.~\citet{walczak2015,fontes2015,wollaeger2018,2020MNRAS.493.4143F,wollaeger2019}.  For the current radiation-hydrodynamics calculations, we assume a solar metallicity distribution of elements~\citep{gs1998} and consider this mixture down to very low densities~\citep{2013ApJS..204...16F}.  

Implementing opacities into a radiation-hydrodynamics code also requires approximations.  Many calculations use a single opacity for the entire frequency space.  How the opacity is weighted to create this opacity is important and depends on the problem~\cite{2007rahy.book.....C}.  In many applications, the Rosseland approximation is used:
\begin{equation}
\kappa^{-1} = \frac{\int_\nu{\kappa_\nu^{-1} \partial{B_\nu(T)}/\partial{T}}}{\int_\nu{\partial{B_\nu(T)}/\partial{T}}}
\end{equation}
where $\kappa_\nu$ is the energy/frequency dependent opacity from our atomic physics calculations and $B_\nu$ is the blackbody photon distribution.  For a single gray (frequency-independent) opacity, this prescription highlights the low-opacity regimes (valleys) where radiation can leak out.  Another extreme is the Planck prescription:
\begin{equation}
\kappa = \frac{\int_\nu{\kappa_\nu B_\nu(T)}}{\int_\nu{B_\nu(T)}}.
\end{equation}
This weights the opacity in the lines where the opacity is largest.  In a free-streaming environment where the radiation is described by a flux at a photosphere ($F_\nu$), the Planck description can be further simplified by using a single flux.  For line-driven effects of optically-thin stellar winds, these flux-weighted opacities can be used~\citep{Sander20}:
\begin{equation}
\kappa = \frac{\int_\nu{\kappa_\nu F_\nu}}{\int_\nu{F_\nu}}.
\end{equation}
Rosseland tends to underestimate the opacity and Planck or Flux-weighted methods overestimate the opacity.  Multi-group approaches capture both the high- and low-opacity regions in energy space.  The differences between the approximate (Rosseland, Planck, flux-weighted) methods decrease with higher group number. With our multi-group prescription, we varied the number of groups to find a convergence in our answer, but found reasonable convergence with our 24-group scheme and use that for the simulation in this paper. To produce detailed spectra, our post-process calculations employ the full 14900-point, monochromatic opacities contained in the OPLIB tables.

This opacity implementation does not include a couple physics issue studied in the supernova literature: NLTE opacities and the expansion opacity corrections.  NLTE opacity effects are strongest as the density decreases and electron collisions no longer dominate the distribution of excited states in an atom.  During shock breakout, electron collisions remain rapid and the errors introduced by our LTE assumption are still minimal.  In addition, while including line-broadening from Doppler effects, our binned approach does not include the full physics required to do a full expansion opacity~\citep{2007rahy.book.....C}.  At high optical depths, the corrections derived from expansion opacity approaches can be very important.  However, at low optical depths, expansion opacity recipes all converge to the line-binned approach~\citep{2020MNRAS.493.4143F}.  Different expansion opacity recipes converge differently and it is difficult to determine which convergence is most accurate.  At modest to low optical depths, the best implementation of the opacity has yet to be determined.  However, tests of these different implementations show that the line-binned approach does give very similar answers to expansion opacity approaches in kilonva and type Ia supernova applications~\cite{2020MNRAS.493.4143F}.

\section{Physics Behind Shock Heating in Shock Breakout}
\label{sec:shockphys}

As the supernova shock propagates through the star, the shock velocity ($v_{\rm shock}$) and position ($r_{\rm shock}$) are well-fit by assuming the Sedov-Taylor similarity solution:
\begin{equation}
r_{\rm shock} \propto t_{\rm shock}^{2/(5-\omega)} (E_{\rm SN}/\rho_0)^{1/(5-\omega)}
\end{equation}
where $E_{\rm SN}$ is the explosion energy, the density of the circumstellar medium is given by $\rho_0 r^{-\omega}$, and $t_{\rm shock}$ is the propagation time.  The corresponding shock velocity is:
\begin{equation}
v_{\rm shock} \propto t_{\rm shock}^{(\omega-3)/(5-\omega)} (E_{\rm SN}/\rho_0)^{1/(5-\omega)}
\label{eq:sedov}
\end{equation}
For a simple wind profile ($\rho \propto r^{-2}$), then shock velocity is simply given as a function of time ($v_{\rm shock} \propto t_{\rm shock}^{-1/3}$) and radius ($v_{\rm shock} \propto r_{\rm shock}^{-1/2}$).  This deceleration will produce a reverse shock that heats the material, increasing the temperature.  As long as the radiation is trapped in the flow, this Sedov-Taylor solution is a good approximation of the shock evolution.

However, as the shock first becomes optically thin, the radiation begins to lead the shock.  In the extreme limit, one can assume that the radiation is just an energy sink (the beginning of the snowplow phase where the shock propagation is simply determined by momentum conservation).  However, at shock breakout, although the radiation is streaming out of the star, it still couples to the circumstellar medium.  This coupling deposits momentum into this medium ahead of the shock.  The acceleration ($d v_{\rm radiation}/dt$) of this material is the same as the assumptions made in the derivation of the Eddington limit:
\begin{equation}
d v_{\rm radiation}/dt = \kappa L_{\rm breakout}/(4 \pi r^2 c)
\end{equation}
where $\kappa$ is the opacity (for hydrogen, $\kappa \approx 0.4 \, {\rm cm}^2 \, {\rm g}^{-1}$), $L_{\rm breakout}$ is the luminosity of the breakout emission, $r$ is the radius and $c$ is the speed of light.  For our shock breakout, we can estimate the acceleration caused by the radiation:
\begin{equation}
\frac{d v_{\rm radiation}}{dt} = 290 \left(\frac{r_{\rm Breakout}}{r}\right)^{2} \left(\frac{T_{\rm shock}}{\rm 40~eV}\right)^4 \left(\frac{\kappa}{0.4 \, {\rm cm}^2 \, {\rm g}^{-1}}\right) {\rm \, km \, s^{-2}}
\label{eq:radacc}
\end{equation}
where $T_{\rm shock}$ is the temperature at the photosphere where the photons are breaking out of the supernova shock, $r_{\rm Breakout}$ is the radius of this photosphere, and $r$ is the position of the material being accelerated.  Material just ahead of breakout could easily be accelerated above the shock velocity just due to this photon momentum deposition (e.g., for $r_{\rm Breakout}/r =1/3$, the shock can accelerate to $10^4~{\rm km \, s^{-1}}$ in roughly 300~s). 

The relative velocity of the supernova blast wave to its circumstellar medium dictates the amount of heating in the shock.  As the radiation-driven shock of the supernova hits a clump or shell, it shocks, both compressing the material and heating it.  The corresponding density ($\rho_{\rm shock}$) and pressure ($P_{\rm shock}$) at the shock in the strong shock limit where the shock velocity is much greater than the speed of sound ($v_{\rm shock} > \sim 100 c_{\rm sound}$ in our supernova calculations) are:
\begin{equation}
    \rho_{\rm shock}=\frac{\gamma+1}{\gamma-1} \rho_{\rm clump}
\end{equation}
and
\begin{equation}
    P_{\rm shock}=\frac{\gamma+1}{2} \rho_{\rm clump} v^2_{\rm shock}
\end{equation}
where $\rho_{\rm clump}$ is the density of the clump and $v_{\rm shock}$ is the shock velocity.  In shock breakout where the radiation is still reasonably coupled to matter, $\gamma \approx 4/3$ and the pressure is roughly the radiation pressure:  $P_{\rm shock} = (a/3) T^4_{\rm shock}$ where $a$ is the radiation constant.  In this scenario, the temperature of the shock depends on the shock velocity and the clump density:
\begin{equation}
    T_{\rm shock}=35~eV (\rho_{\rm clump}/10^{-10} {\rm \, g \, cm^{-3}})^{1/4} (v_{\rm shock}/10^9 {\rm \, cm \, s^{-2}})^{1/2}
    \label{eq:sjc}
\end{equation}

The strong shock limit is an oversimplification of the physics in this problem.  For example, the energy in the supernova blastwave can, in some instances, be on par with the energy injected in the shock. In such a scenario, the shock pressure is closer to:
\begin{equation}
    P_{\rm shock}=\frac{\gamma+1}{2} \rho_{\rm clump} v^2_{\rm shock} + (a/3) T_{\rm Blastwave}^4
\end{equation}
where $T_{\rm Blastwave}$ is the temperature of this blastwave.  For a 40-eV blastwave, $\rho=3\times10^{10} {\rm \, g \, cm^{-3}}$ density shell, we expect a jump in temperature of $\approx 10 {\rm \, eV}$ up to roughly ${\rm 50\,eV}$.  If the density is 10 times lower, the jump should be 1--2\,eV. For an equivalent $10\,{\rm eV}$ blastwave, we expect a jump in temperature of $\approx 15 {\rm \, eV}$ up to roughly 25--30 eV.  In addition, the supernova shock can flow around clumpy material rather than produce the strong shock conditions assumed in equation~\ref{eq:sjc}.  Radiative acceleration of the circumstellar medium can limit the amount of shock heating because it is the relative velocity of the shock and the circumstellar medium that dictates the heating.  Simulations are required to move beyond simple analytic estimates of shocks in shock breakout.

A number of laboratory experiments have been developed to test radiation flow physics in conditions are relevant to supernova breakout conditions~\citep[e.g.,][]{2015JQSRT.159...19M,2016HEDP...18...45F,2018NatCo...9.1564K}.  In particular, the Radishock experiment (Wood, S., et al. in preparation) is designed to mimic the conditions of shocks in shock breakout conditions.  This experiment drives a target on two sides, a radiation-driven shock on one end and a matter driven shock on the other.  The impact of these two shocks mimics the shock interactions in shock breakout and these experiments can be used to understand this physics.  These experiments have been used to test the codes used in the calculations presented here.

\section{Radiation Hydrodynamics Simulations}
\label{sec:radhydro}

For this paper, we have developed a simplified shock breakout scenario, focusing on the role inhomogeneities can play on the emission from shock breakout.    Our supernova shock is implemented using an inflow boundary condition driving a radiative shock through a wind density profile.  This inflow boundary condition is characterized by a shock temperature (which we vary), a shock velocity, and density.  The temperature is held fixed for 500\,s\footnote{This dictates the structure behind our shock and although it is arbitrary, the shock interactions play a much bigger role in dictating both the evolution of the blastwave and the breakout emission.} and then is allowed to decrease.

In many stars, shock breakout occurs within the transition between the "stellar edge" and a constant-velocity wind profile.  In this region, the wind velocity increases outward.  This transition region is dictated by the radiative acceleration.  Typically, the velocity of the wind reaches an asymptotic limit and, beyond this limit, the velocity can be assumed to be constant.  This constant-velocity limit is not fully reached in the transport regime modeled in this study but, for the purposes of studying the role of inhomogeneities in the shock breakout region, a simple wind profile assuming a constant velocity, constant mass-loss wind provides a 
background to compare smooth and inhomogeneous density distributions.  These assumptions produce a $r^{-2}$ density ($\rho_{\rm wind}$) profile:
\begin{equation}
    \rho_{\rm wind} = \dot{M}_{\rm wind}/(4 \pi r^2 v_{\rm wind})
\end{equation}
where $\dot{M}_{\rm wind}$ is the wind mass-loss rate, $v_{\rm wind}$ is the wind velocity (we use a constant $10^8 \, {\rm cm s^{-1}}$ for this paper) and $r$ is the radius from the star.  These outbursts include a wide range of instabilities.  For example, non-spherical oscillations in convective shell burning can cause explosive burning that drives mass ejection in a stellar model~\citep{2014ApJ...792L...3H}.  The ejecta from these outbursts is highly asymmetric.  Alternatively, binary systems can undergo a common envelope phase that both ejects matter and alters the structure of the star~\citep{2012ApJ...744...52P,2013A&ARv..21...59I,2020arXiv200306151I}.  These systems also have highly asymmetric outflows as well.  A final example is the inhomogeneous outflows from line-driven winds~\citep{1984ApJ...284..337O, 2008A&ARv..16..209P, 2018Natur.561..498J,2019MNRAS.485..988O}.  We study two suites of simplified instantiations of inhomogeneities in shock breakout:  spherical clumps mimicking turbulent instabilities in the wind and dynamic ejecta mimicking stellar outbursts prior to the launch of the supernova explosion.  A summary of the set of calculations are listed in table~\ref{tab:models}.

\begin{table}
  \centering
  \scriptsize
  \begin{tabular}{lccc}
  \hline
  \hline
Model & $T_{\rm SN}$ & $\dot{M}_{\rm wind}$ & $V_{\rm shock}$ \\
& (eV) & ($3 \times \rm 10^{-4} M_\odot s^{-1}$) & ($10^4 {\rm km \, s^{-1}}$) \\
\hline
NCT40 & 40 & 1 & 1 \\
NCT20 & 40 & 1 & 1 \\
NCT10 & 40 & 1 & 1 \\
\hline
CT40M1 & 40 & 1 & 1 \\
CT40M1v2 & 40 & 1 & 2 \\
CT40M1ext & 40 & 1 & 1 \\
CT40M1extv2 & 40 & 1 & 2 \\
CT20M1 & 20 & 1 & 1 \\
CT10M1 & 10 & 1 & 1 \\
CT40M0.1 & 40 & 0.1 & 1 \\
CT40M0.1v2 & 40 & 0.1 & 2 \\
CT40M0.1extv2 & 40 & 0.1 & 2 \\
CT20M0.1 & 20 & 0.1 & 1 \\

\hline
\hline

Model & $T_{\rm SN}$ & $\dot{M}_{\rm wind}$ & $r_{\rm shell}$ \\
& (eV) & ($3 \times \rm 10^{-4} M_\odot s^{-1}$) &  $10^{12} {\rm \, cm}$ \\
\hline
sh0.1r5T40 & 40 & 0.1 & 0.5 \\
sh0.1r1T40 & 40 & 0.1 & 1 \\
sh0.1r2T40 & 40 & 0.1 & 2 \\
sh1r5T40 & 40 & 1 & 0.5 \\
sh1r1T40 & 40 & 1 & 1 \\
sh1r2T40 & 40 & 1 & 2 \\
sh1r5T10 & 10 & 1 & 0.5 \\

  \end{tabular}
  \caption{Models:  smooth winds, clumps and shells.  The 'NC' models are smooth wind profiles.  The `ext' models have clumps that extend to $5\times10^{13} {\rm cm}$.}
  \label{tab:models}
\end{table}

Our 2-dimensional calculations model a long, narrow ($8\times10^{11} \, {\rm cm}$) slice of the circumstellar medium, modeling from the edge of a Wolf-Rayet star at $10^{11} {\rm cm}$ out to $6.4\times10^{13} \, {\rm cm}$.  Our base resolution is $2\times10^{10} \, {\rm cm}$ and we only allow for one additional level of refinement (fine resolution of $10^{10} \, {\rm cm}$).  

\subsection{Shock Interactions with Shells}

We mimic the case where activity on the surface of the star ejects a shell of material by introducing a constant density slab of material.  We vary the position of this shell and give it a thickness equal to its position, i.e. a shell of material at $5\times10^{11} {\rm \, cm}$ extends to $10^{12} {\rm \, cm}$, a shell starting at $10^{12} {\rm \, cm}$ extends to $2 \times 10^{12} {\rm \, cm}$.  The shell is assumed to have a constant density that is 10 times that of the wind density at the inner position of the shell. 

\begin{figure}[ht!]
\plotone{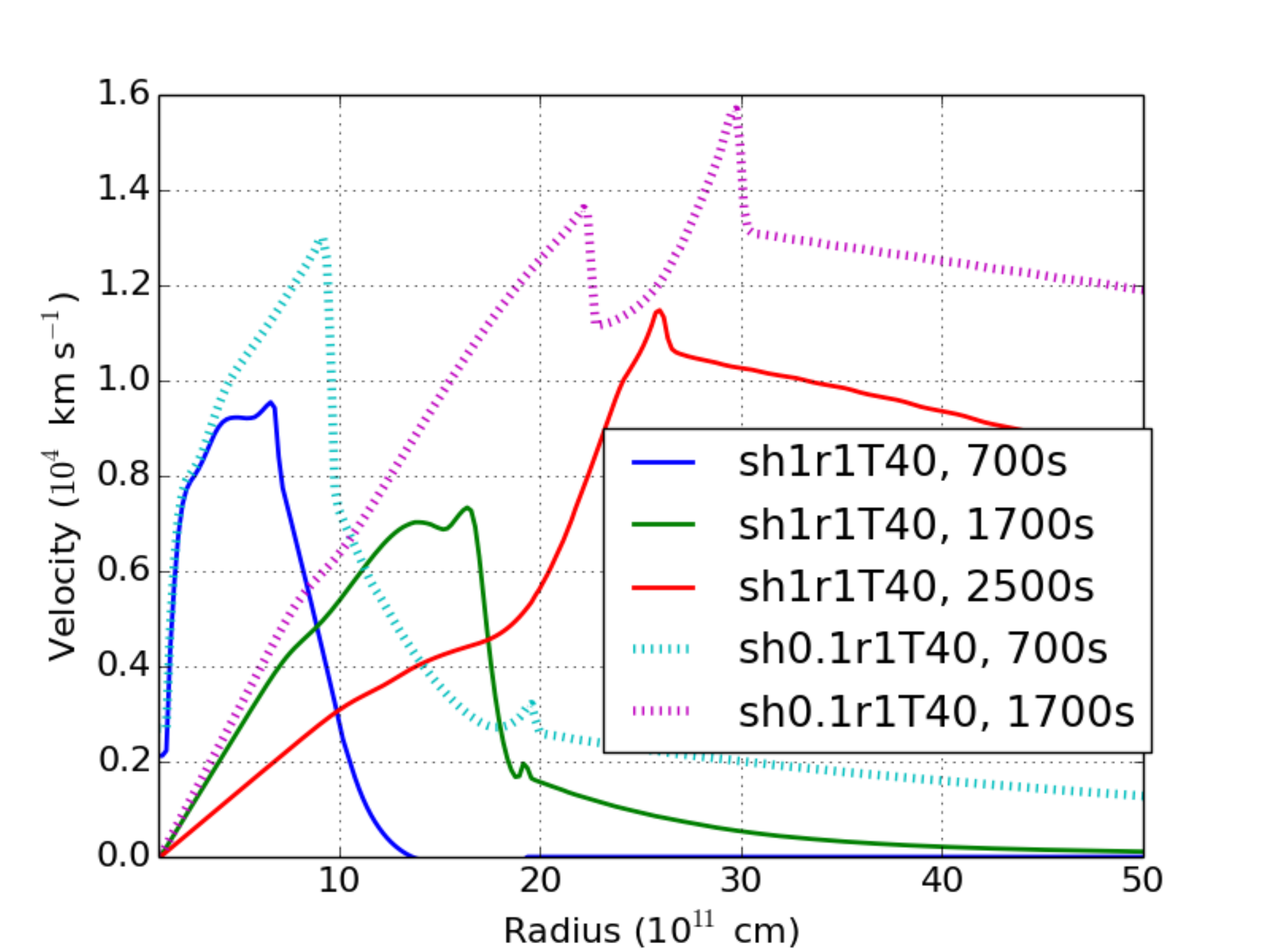}
\caption{Velocity versus radius for 2 supernova models, both with shells placed at $10^{12} {\rm \, cm}$ at a few times.  The solid lines correspond to a model with a wind mass loss of $3\times10^{-4} {\rm \, M_\odot s^{-1}}$ and the dotted lines correspond to a model with a mass loss of  $3\times10^{-5} {\rm \, M_\odot s^{-1}}$.  Radiation-driven shocks become important for the low mass-loss case at roughly 500~s and at roughly 2000~s for the high mass-loss case.  The radiation-trapped flow is well matched by a Sedov-Taylor solution (equation~\ref{eq:sedov}) and the radiatively-driven flow is matched by our radiative solution (equation~\ref{eq:radacc}).}
\label{fig:shellvel}
\end{figure}

Figure~\ref{fig:shellvel} shows the velocity profile of the supernova blastwave for 2 different wind densities (assuming mass loss rates of $3\times10^{-5}, 3\times10^{-4} {\rm \, M_\odot s^{-1}}$ for shell models at $10^{12} {\rm \, cm}$.  These two models show both extremes of our velocity evolution.  For our high-density wind model where the radiation is initially trapped, the velocity evolution is very close to the Sedov-Taylor solution.  For example, the blastwave velocity drops by nearly 30\% (our simple solution predicts a 40\% decrease) between 700 to 1700s.  At later times, the radiation begins to lead the blastwave, accelerating the material ahead of it.  If we assume the accelerating region is 3 times the shock position, using our radiative formula (equation~\ref{eq:radacc}), we'd expect the shock to reach velocities of $15,000 {\rm \, km \, s^{-1}}$ after 500\,s.  This is an upper limit on the acceleration because, once the radiatively-accelerated shock forms, it moves away from the initial blastwave, reducing its acceleration.  It will also lose momentum by emitting photons itself.  The velocity of the radiatively-driven shock for this model rises to $12,000 {\rm \, km \, s^{-1}}$ in less than a 800~s timescale.  The radiation escapes from the low-density wind model in the first 700~s, producing a strong radiatively-driven shock.

\begin{figure}[ht!]
\plotone{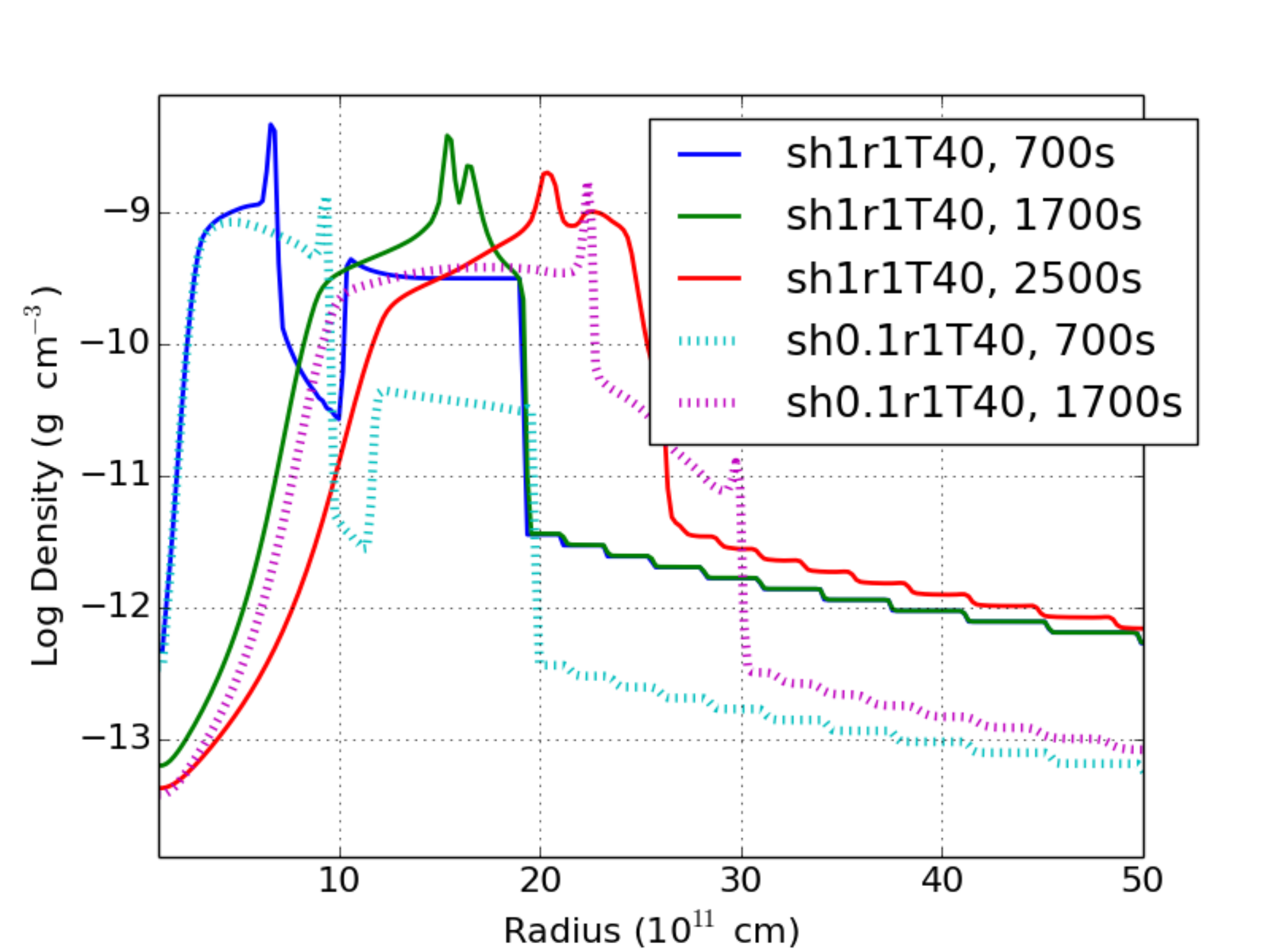}
\plotone{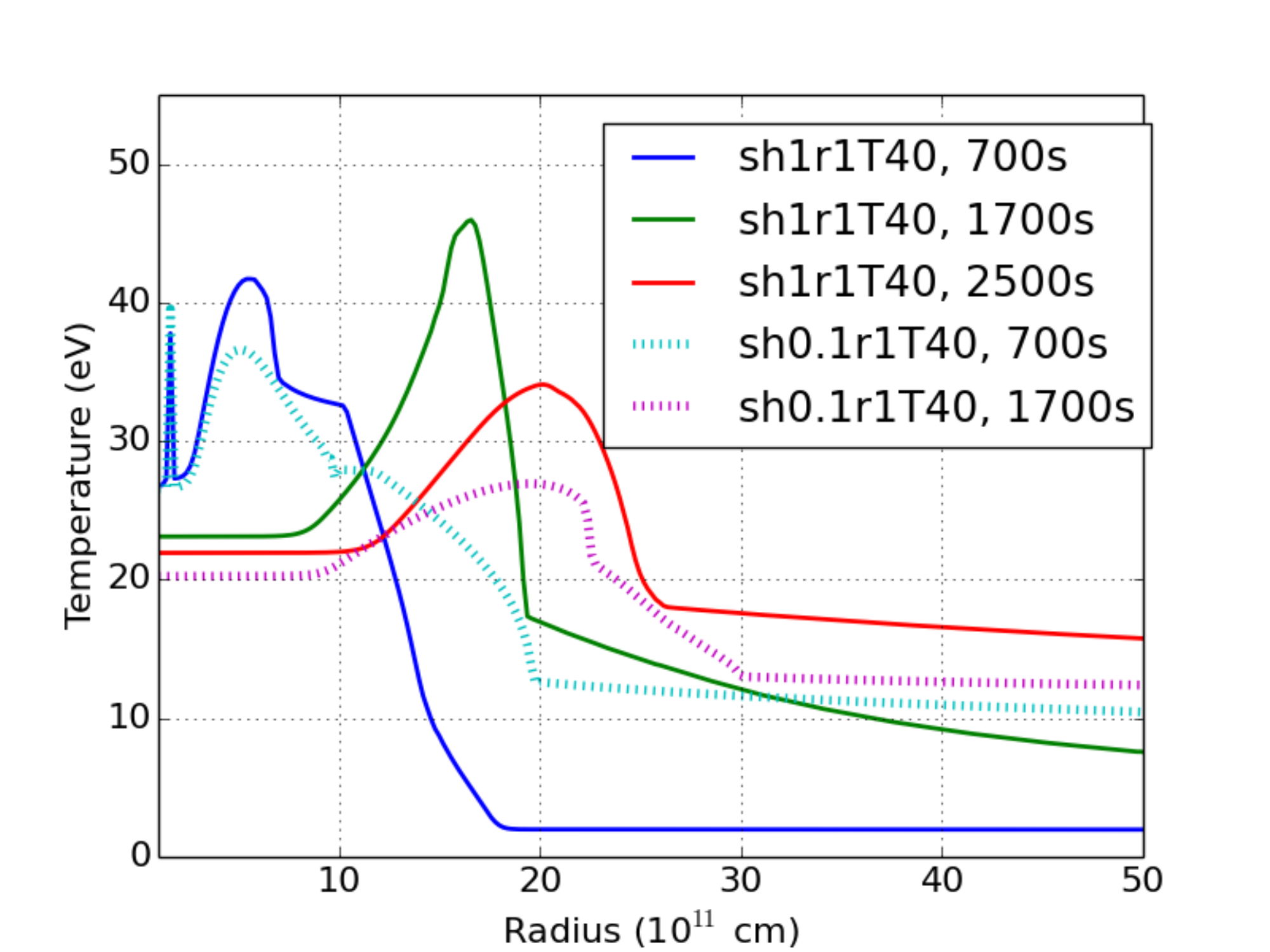}
\caption{Shell density (top) and temperature (bottom) as a function of time for the same 2 basic models in Figure~\ref{fig:shellvel}.  Although our simple shock-heating model fits well the trends and rough values of the peak temperatures, radiative acceleration and heating make it difficult to produce exact analytic fits.   \label{fig:shelldentemp}}
\end{figure}

Because of this radiatively-driven shock both contributing to the acceleration and heating of the shock, it is much more difficult to make accurate analytic temperature and density estimates of the supernova blast.  Figure~\ref{fig:shelldentemp} shows the density and temperature profiles of our shell models from Figure~\ref{fig:shellvel}.  If the radiation is trapped, our shock heating derivation predicts a peak temperature that increases by a few to 5\,eV higher than our 40\,eV drive, in agreement with our simulations.  However, once the radiation is no longer trapped, the shock front cools considerably.  Even for the situation where the radiation is not trapped, there is a jump in the shock density (on par with the strong shock estimate) and temperature (of 3--10\,eV) as the blastwave hits the shell.  These values are within the order of magnitude expectations from our analytic derivation.

\begin{figure}[ht!]
\plotone{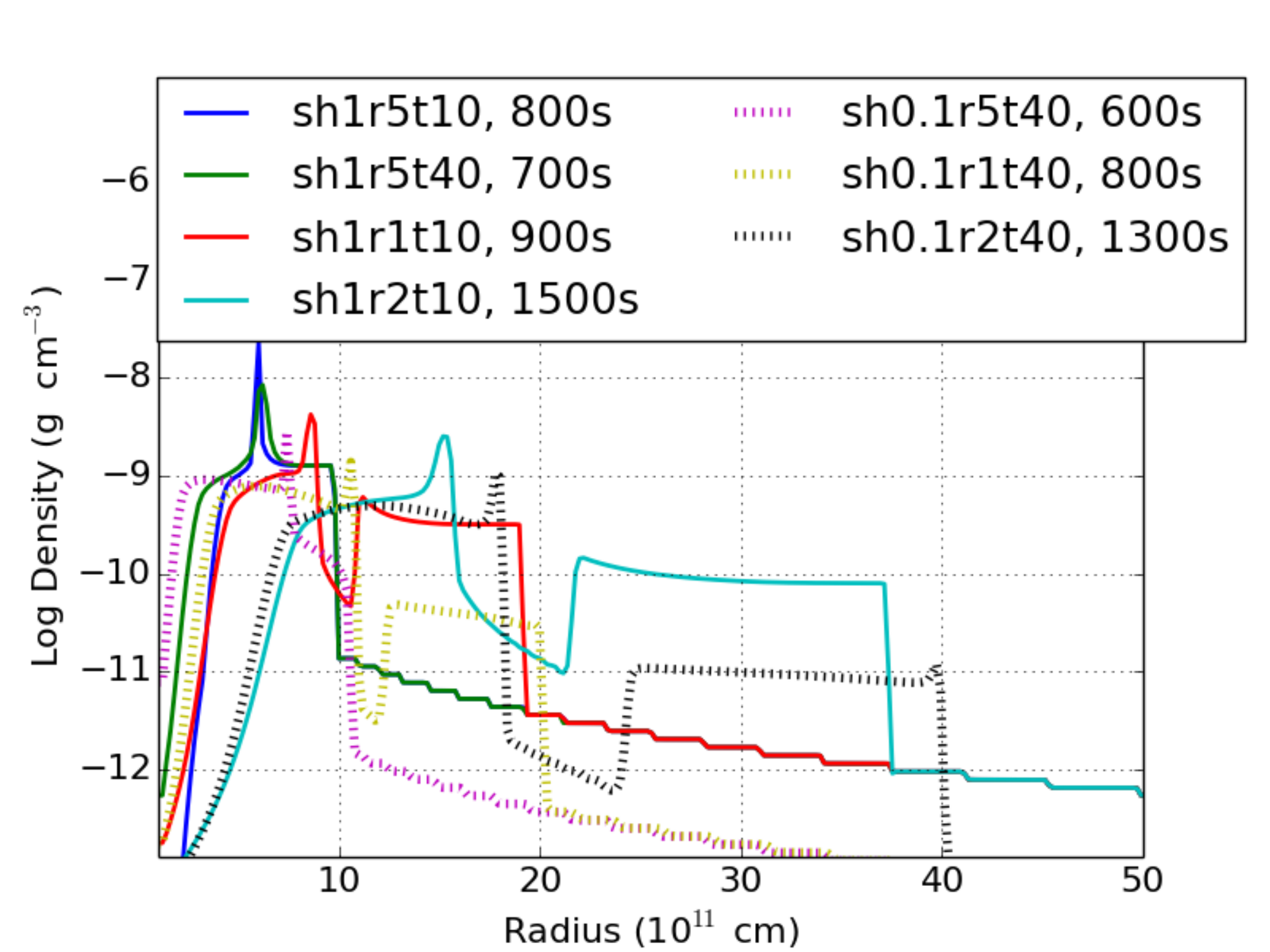}
\plotone{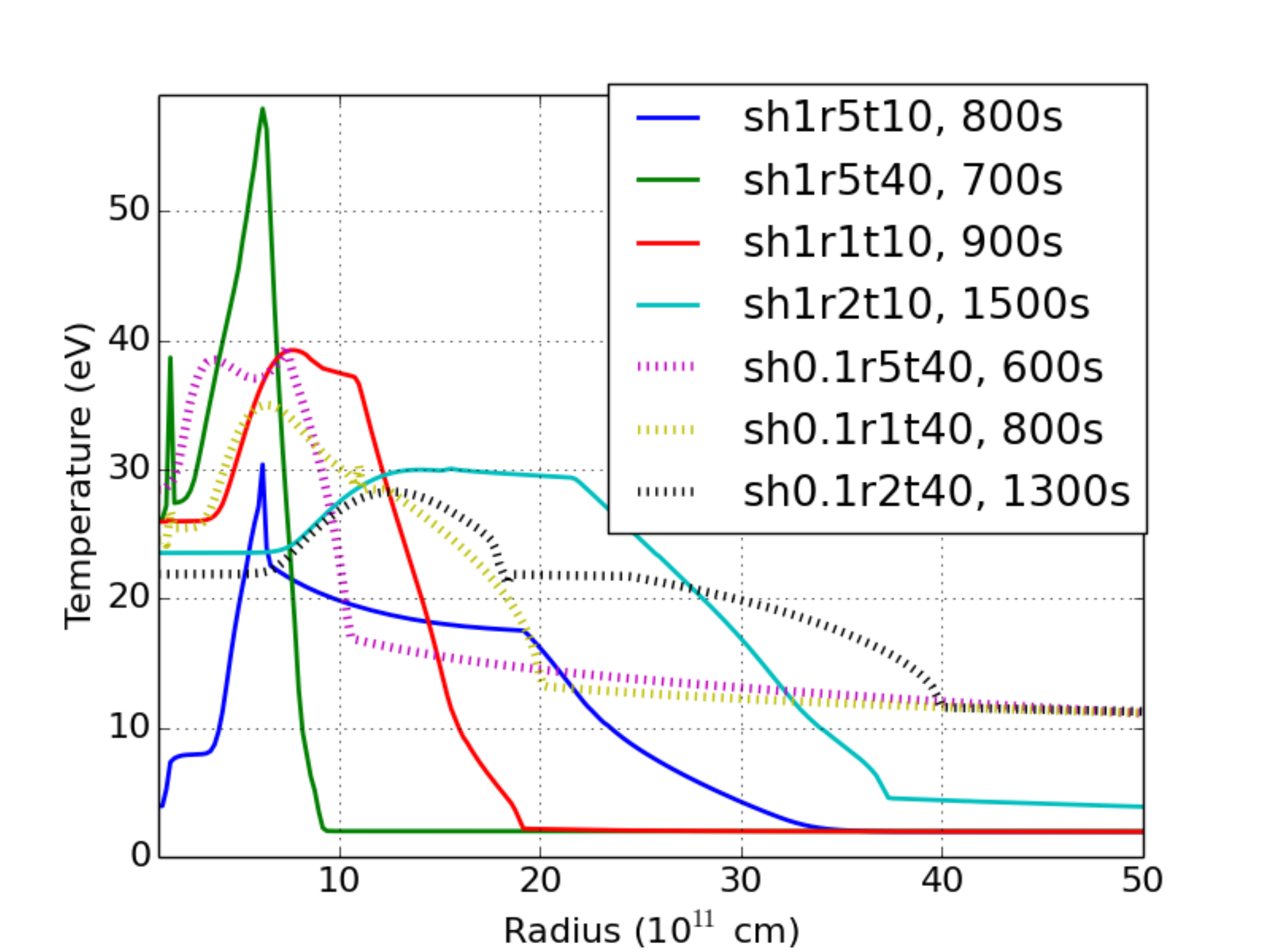}
\caption{Density (top) and temperature (bottom) comparisons of many of our shell models at peak temperature.  Although our simple shock-heating model fits well the trends and rough values of the peak temperatures, radiative acceleration and heating make it difficult to produce exact analytic fits.}
\label{fig:shellpeak}
\end{figure}

The peak temperatures for all of our models after shock interactions, along with the densities at those peak temperatures are shown in Figure~\ref{fig:shellpeak}.  Here we see the broad range of results that we can expect from different shell positions (Figure ~\ref{sec:shockphys}).  The peak temperature depends on the position of the shock and the density of the wind as well as the wind density and the position of the shells.  Shell interactions raise the temperature from a few to 20\,eV.  The further out the shell is, the later the peak temperature as it takes longer for the shock interaction with the shell to occur.

\subsection{Shock Interactions with a Clumpy Wind Medium}

For our clumpy wind medium, we introduce 40 randomly-located spheres lying between $10^{11}\, {\rm cm}$ and $5.1 \times 10^{12} \, {\rm cm}$\footnote{We have also included models where the extent of the spheres is out to $5 \times 10^{13} \, {\rm cm}$.  These models have an `ext' in their name.}.  The density of these spheres is randomly set between 1 and 50 of the wind density at the position of the clump (The value of 1-50 is the multiplicative factor that the clump's density is altered to be more dense than the ambient medium).  The radius of the sphere is randomly set to a value between $2\times10^{10}-5\times10^{10} {\rm \, cm}$.  With these "clumps", we can conduct a preliminary study of the shocks produced as the supernova front passes through an inhomogeneous medium.  

\begin{figure}[ht!]
\includegraphics[scale=0.45,angle=-90]{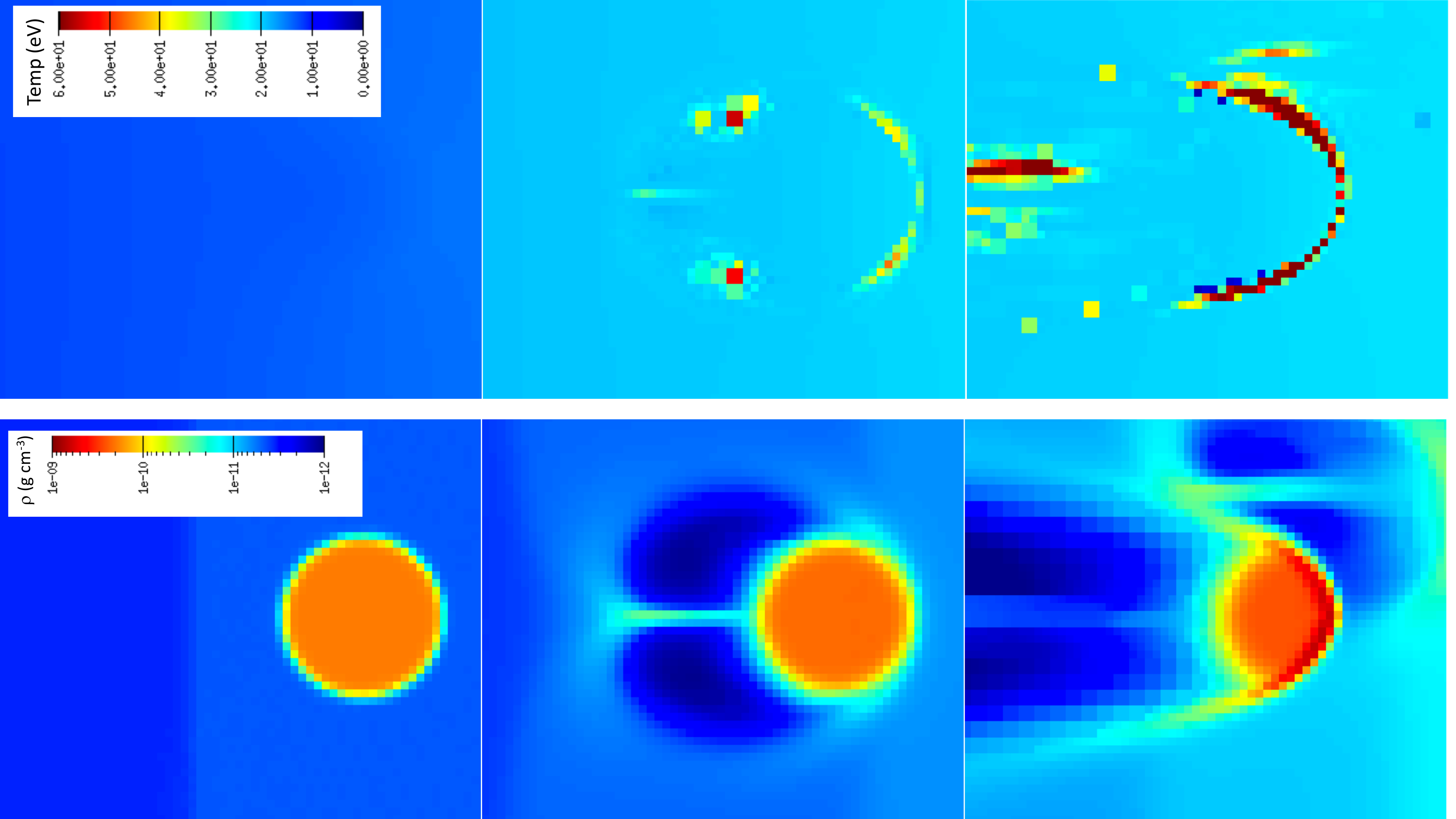}
\caption{Density and temperature of our supernova blastwave passing across just one of our clumps.  The box is $2\times10^{12} {\rm cm}$ by $2.5\times10^{12}{\rm cm}$ and the times from top to bottom are 400, 800, and 1200s after the launch of the shock (this particular clump is near the launch of the shock).  This shows the typical initial behavior of the shock progression.  Material flows around the clump, but the clump shields the downwind material, producing a slightly lower density region.  The front of the clump is shocked, with temperatures slightly lower than that predicted by a strong shock.  But there is also heating behind the clump where the flow interacts.  The lower density in this downwind region allows higher temperatures.}
\label{fig:singleclump}
\end{figure}

To understand the effects of these inhomogeneities, we first zoom into a single clump to better understand its initial evolution.  Figure~\ref{fig:singleclump} shows the density and temperature of this clump shortly after the supernova shock has passed over it.  The shock flows around the clump shocking its sides.  The clump initially shields the region behind it from the shock, producing a low density region that, when forced together, also shocks.  With time, this entire clump is shredded by the flow.  How quickly such a shell is shredded in simulations can depend upon the transport scheme used in a code\footnote{The clumps are not really spheres, but the rapidity at which the clumps are shredded argues that the exact nature of the clumps is not too important.}.  For example, a flux-limited diffusion calculation will quickly heat the clump, causing it to dissipate rapidly.  With higher-order transport schemes, the heating of the clump takes longer.  We have compared (Fryer et al., submitted to HEDP) our $S_N$ method to an implicit Monte Carlo transport method implemented in the same RAGE code framework and, although there are differences, the effect is minimal.  In addition, it is believed that the $S_N$ is more accurate (Fryer et al., submitted to HEDP).  However, continued testing of this physics will ultimately play a role in producing precision results.

\begin{figure}[ht!]
\plottwo{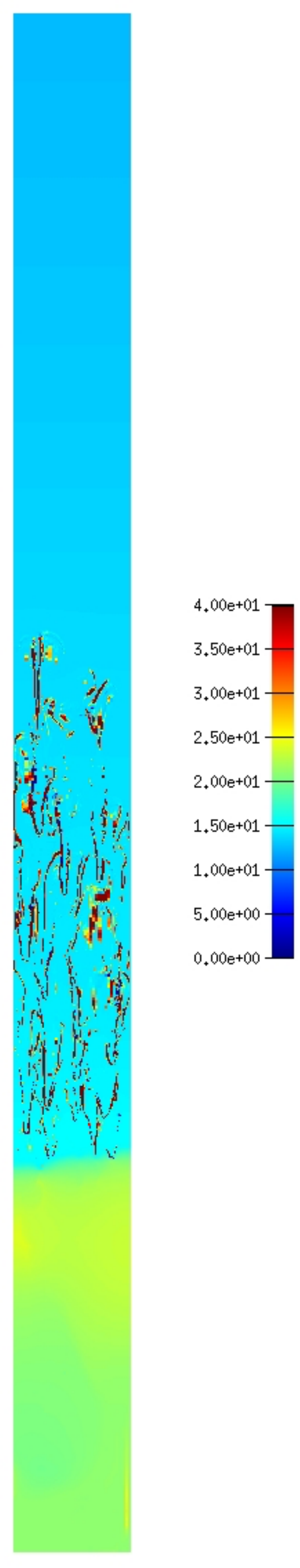}{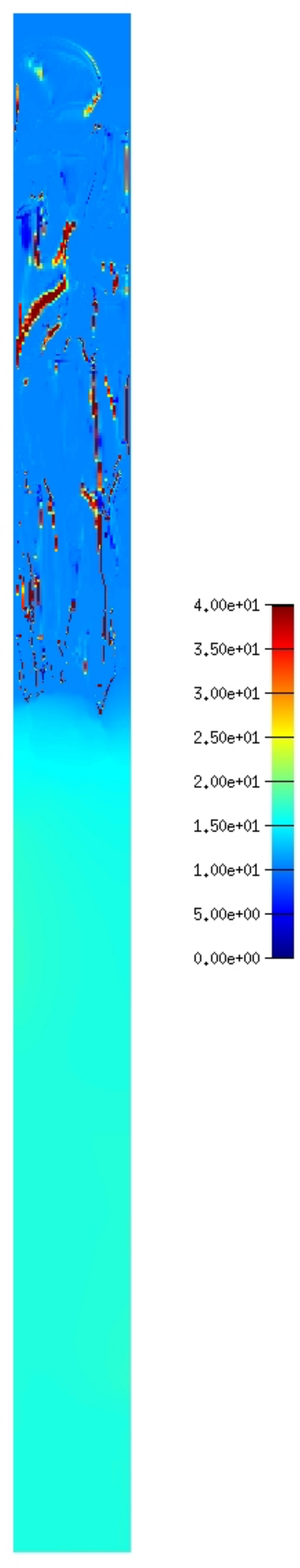}
\caption{Temperature image (eV) of our CT40M1 at 2500 and 5000~s showing the evolution of the initially spherical clumps.  The plot extends from our inner boundary at $10^{11}\, {\rm cm}$ to $10^{13} \, {\rm cm}$.  As the shock progresses, the clumps are shredded, but they leave behind hot spots in the ejecta that will dominate the shock breakout emission.}
\label{fig:clumps}
\end{figure}

Figure~\ref{fig:clumps} shows the progression of our supernova shock (temperature) through the inner region of our simulation (from $10^{11}$ to $10^{13} \, {\rm cm}$).  The supernova shock ultimately shreds the clumps, producing hot tendrils that continue to shock, producing hot, emitting ejecta.  These hot tendrils will emit long-lived UV and X-ray spectra during the breakout signal.

\begin{figure}[ht!]
\plotone{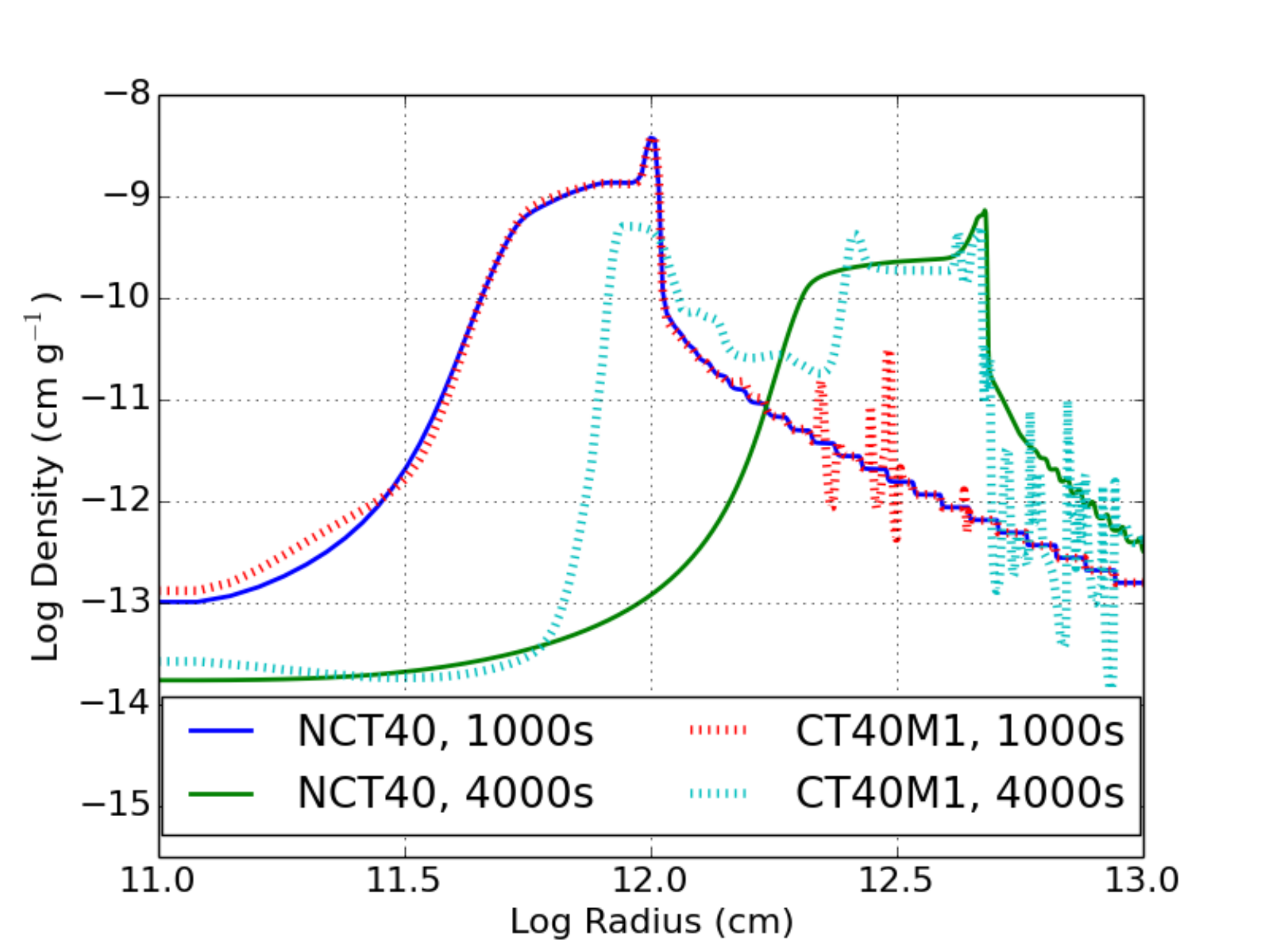}
\plotone{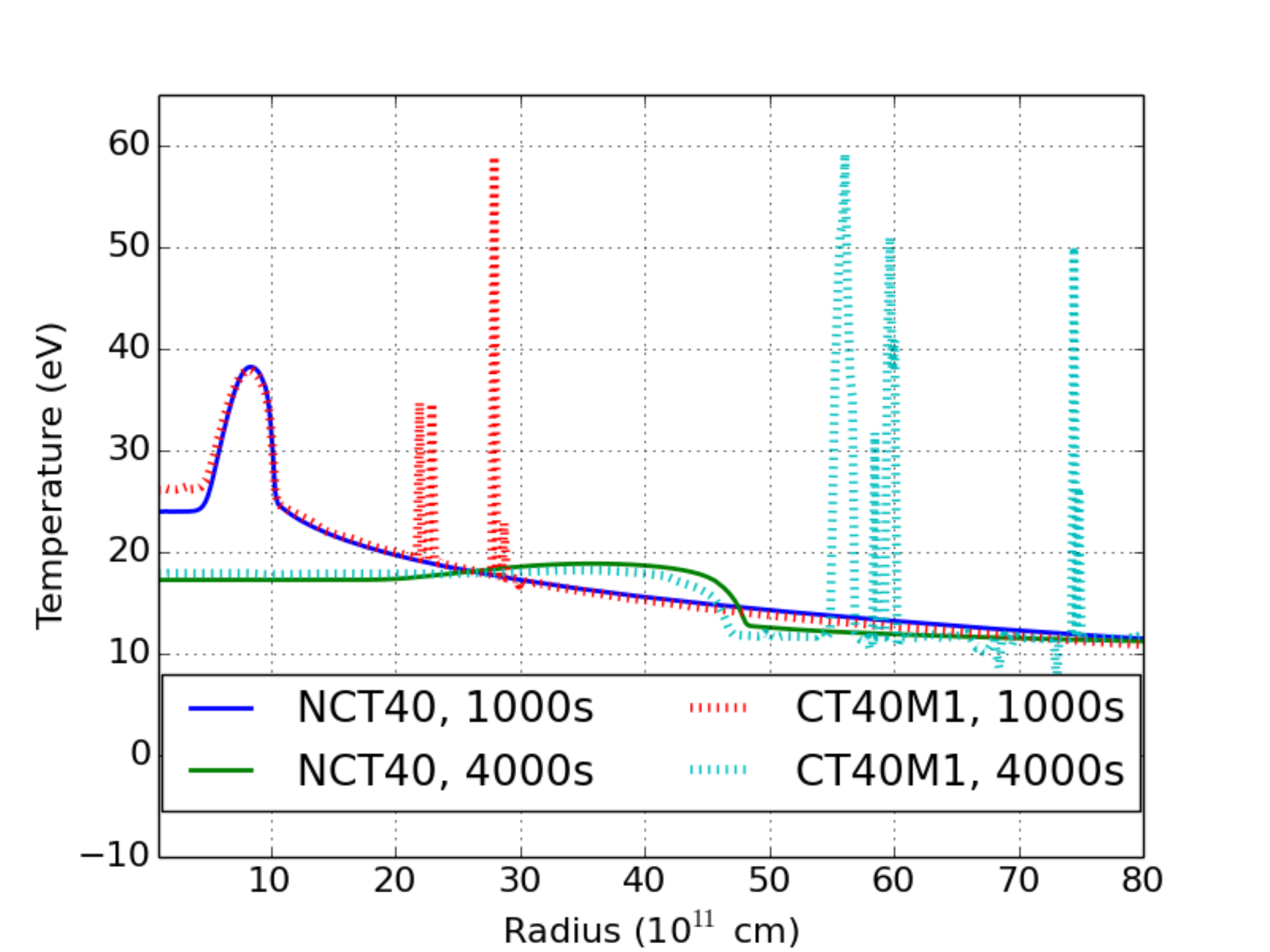}
\caption{Single line-outs of the density (top) and temperature (bottom) of the CT40M1 clump and NCT40 no-clump models at two different times, comparing the simple wind results (solid lines) to the clump results (dotted lines) at two different times.  The clumps lead to structure in both the density and temperature profiles that differ significantly from the simple wind model.  These differences grow with time and will dramatically effect the shock breakout emission in these models. }
\label{fig:clumpprof}
\end{figure}

Figure~\ref{fig:clumpprof} compares the density and temperature profiles as a function of time between the clumpy and simple wind models.  The variable shocks in the clumpy models produce much more variable densities and temperatures, leading to higher peak temperatures that will dramatically alter the emission from these models.  The higher temperatures will produce brighter breakout signals that peak at higher photon energies.  We will study this effect in detail in Section~\ref{sec:spectra}.

\section{Spectra and Light-Curves}
\label{sec:spectra}

The shock heating in these models can signifcantly alter the spectra and light-curves from shock breakout.  Here we review the spectra and light-curves from a few of the models.  

\begin{figure}[ht!]
\plotone{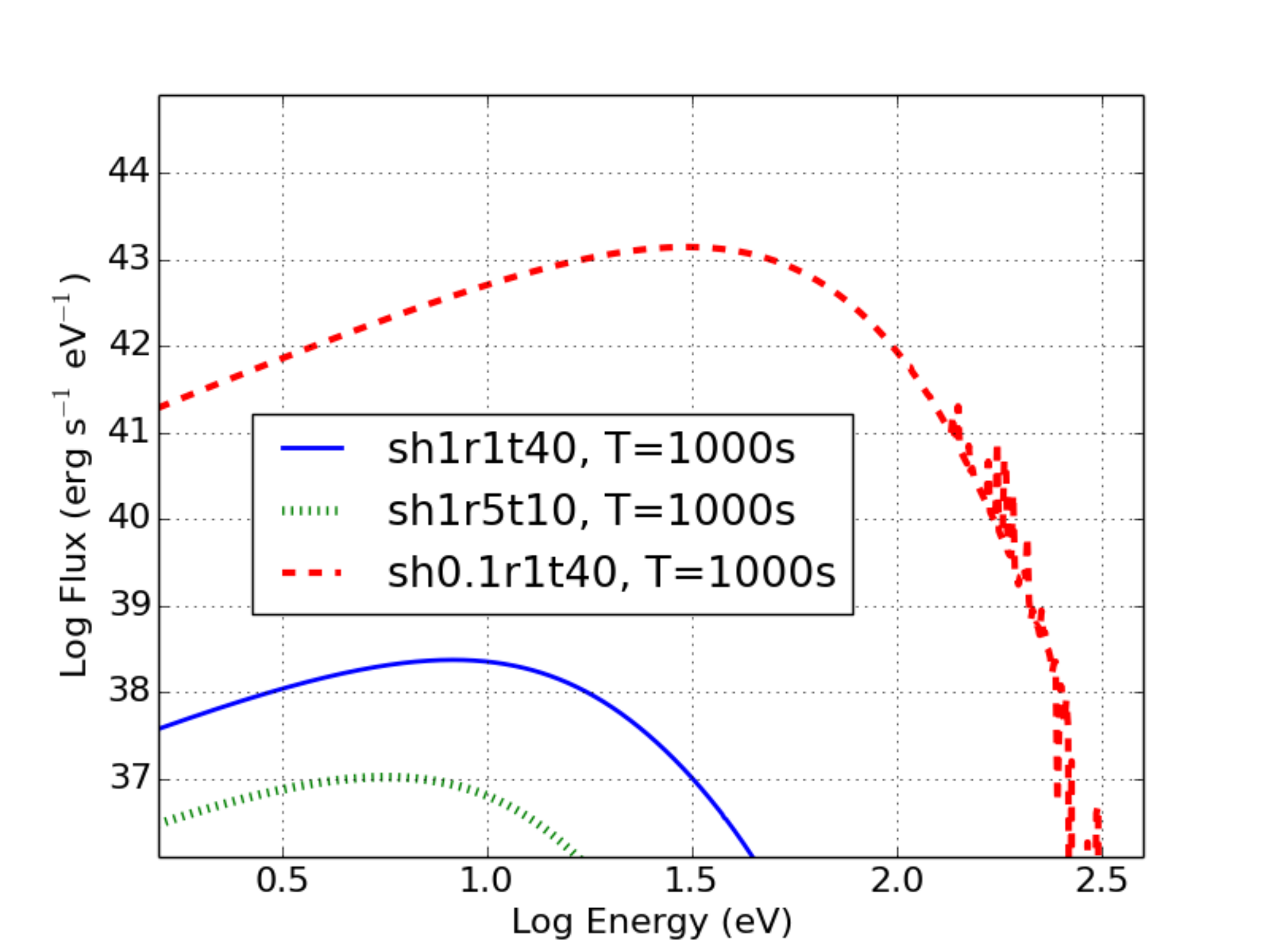}
\plotone{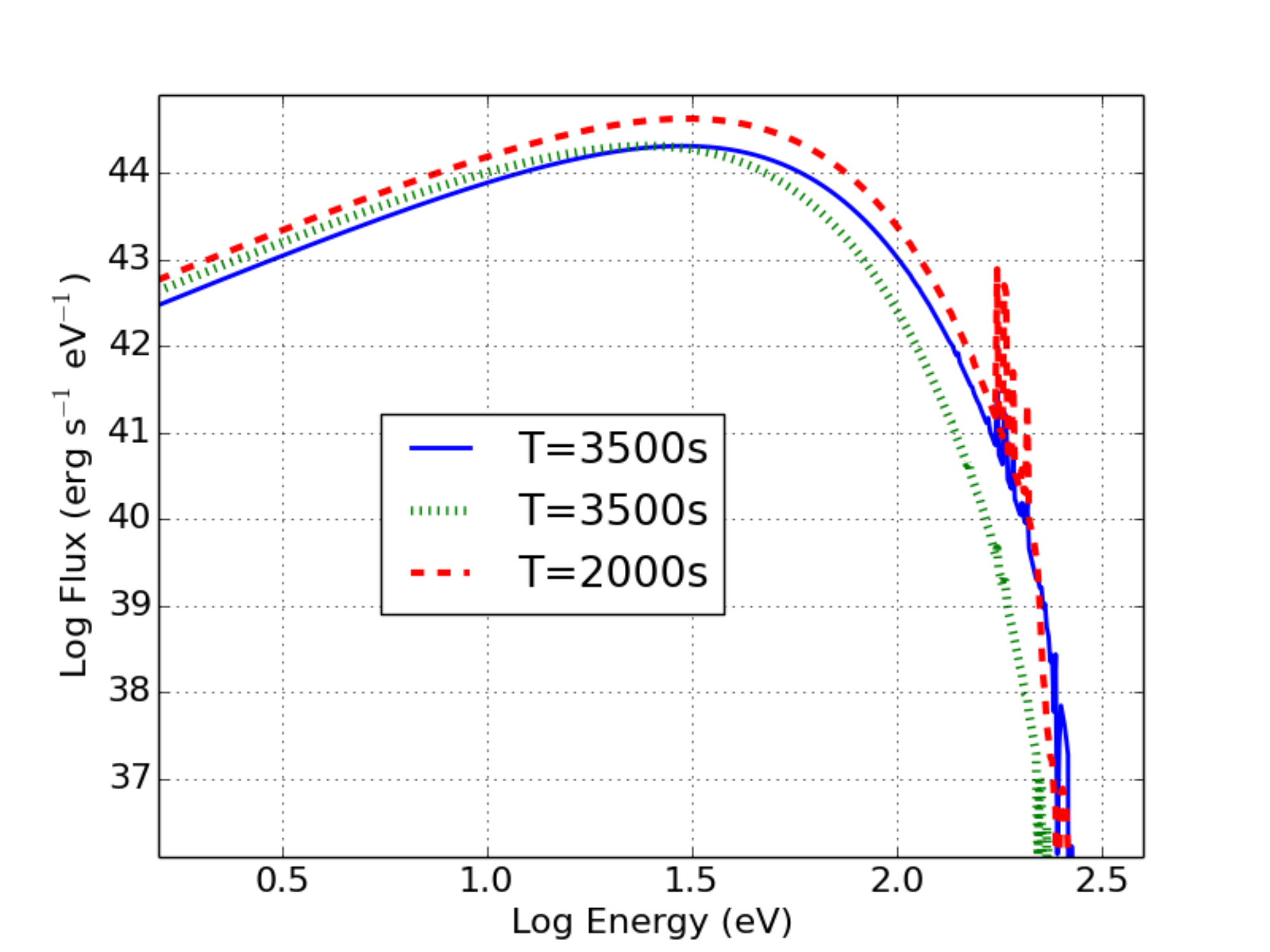}
\plotone{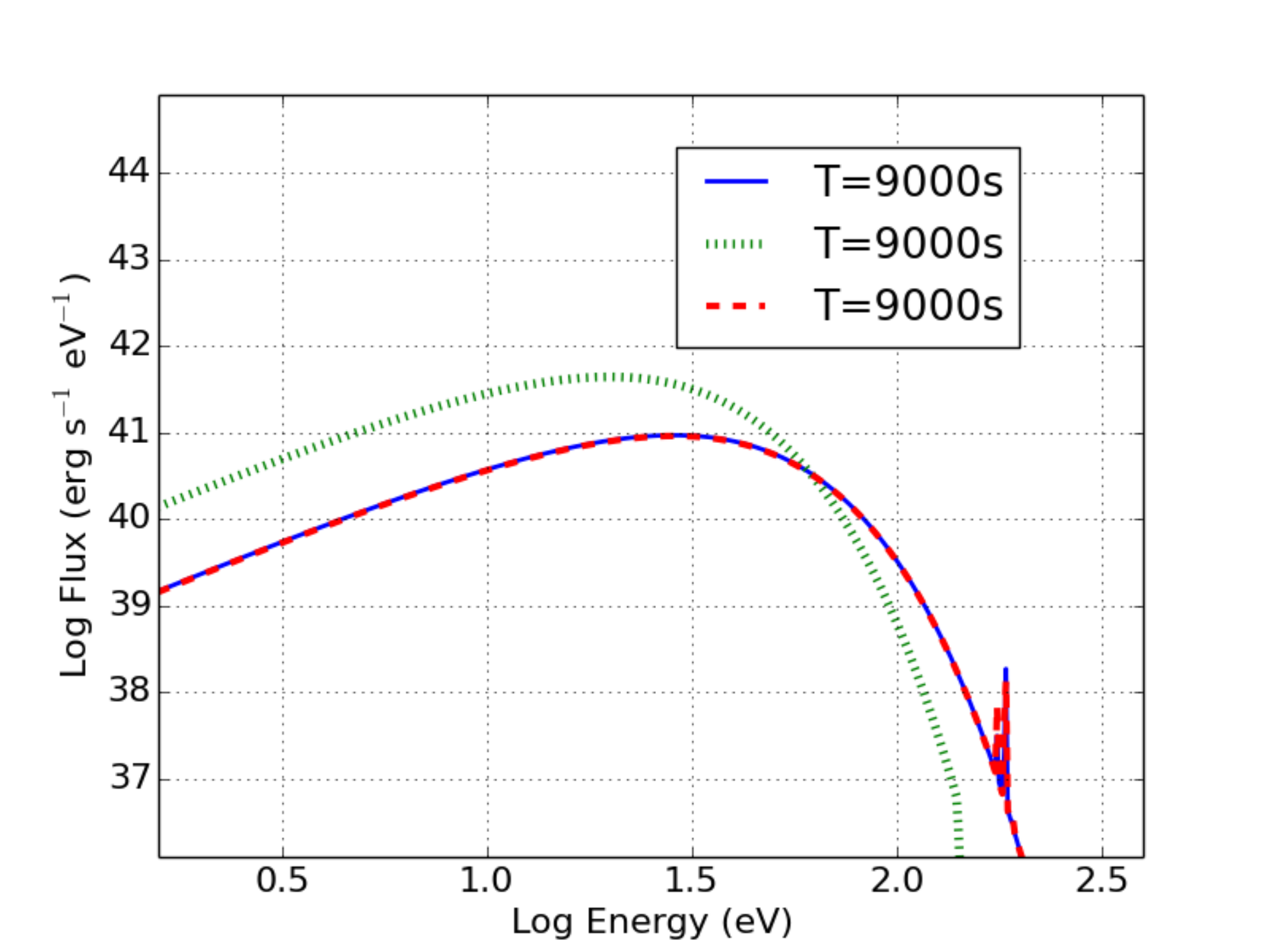}
\caption{Spectra at 1000s (top), peak emission (middle) and at 9000s (bottom) of shell models: sh1r1t40, sh1r5t10, and sh0.1r1t40.  The lower-density wind model rises and peaks earlier than the high-density wind models.  The lower temperature drive model peaks at lower energies.  Nonetheless, the shock heating dominates the temperature of this low density drive, allowing it to produce high energy photons.}
\label{fig:specshell}
\end{figure}

Figure~\ref{fig:specshell} shows a time sequence of spectra for models sh1r1t40, sh1r5t10, and sh0.1r1t40 before peak, at peak and past peak flux.  The low density wind model peaks much more quickly than the two high density models (compare sh0.1r1t40 to sh1r1t40).  Even at peak, it is slightly brighter than the high wind model with the same shell position and drive temperature.  But, at late times, these two models look very similar.  The low temperature drive rises the slowest (the slower drive means that it expands slightly slower) and peaks at a lower temperature, with a lower flux above $\sim 40 \, {\rm eV}$.  The differences are even greater at late times, clearly showing its peak at lower photon energies. 

\begin{figure}[ht!]
\plotone{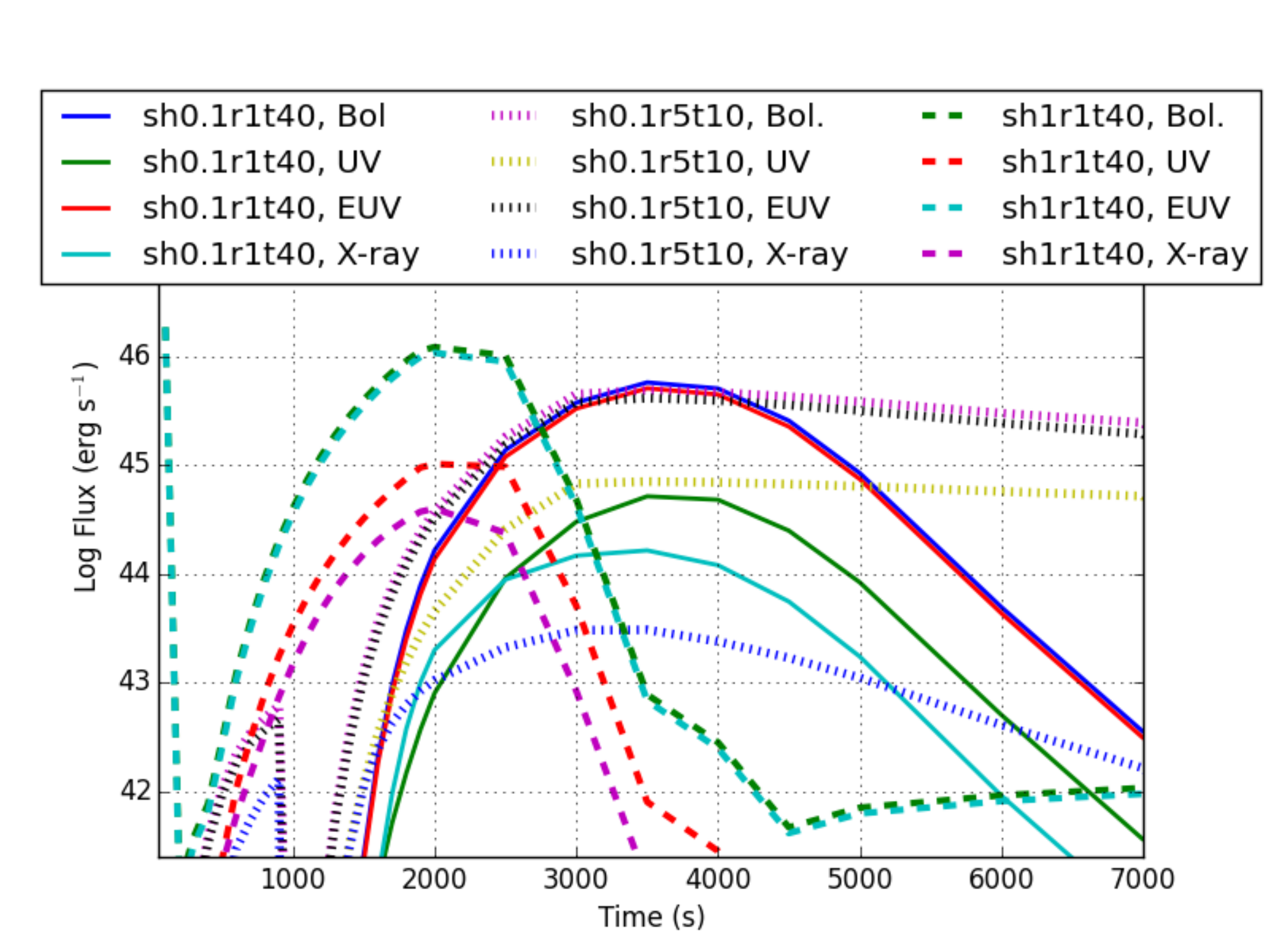}
\caption{Bolometric, UV, extreme UV, and X-ray luminosities as a function of time for 3 of our shell explosion models:  sh11r1t40, sh11r5t10, and sh12r1t40.  The low-density wind case peaks early ($\sim 2000 \, {\rm s}$) versus the roughly 3500\,s peak in our higher-density wind models.  The slower shock in the low temperature drive, close shell model leads to a much slower decay in the emission.  But this low-temperature model produces much less thermal X-ray emission.}
\label{fig:lumshell}
\end{figure}

The corresponding light curves for these 3 models are shown in Figure~\ref{fig:lumshell}.  As expected due to its lower optical depth, the low-density wind model peaks much earlier ($\sim 2000 \, {\rm s}$) than our high-density models ($\sim 3500 \, {\rm s}$).  It is also expected that the low temperature drive model should have a much lower X-ray flux than the other models as its temperature is too low to produce large amounts of X-rays.  What is somewhat surprising is the slow decay for this model.  This occurs both because of the lower drive temperature and the closer shell position.  The high mass of this shell decreases the velocity, producing a slower-moving shock that has a plateau phase before ultimately decaying.  Table~\ref{tab:results} shows  the peak UV and X-ray luminosities, peak luminosity timescale and width of the light-curves for all of our models.

\begin{figure}[ht!]
\plotone{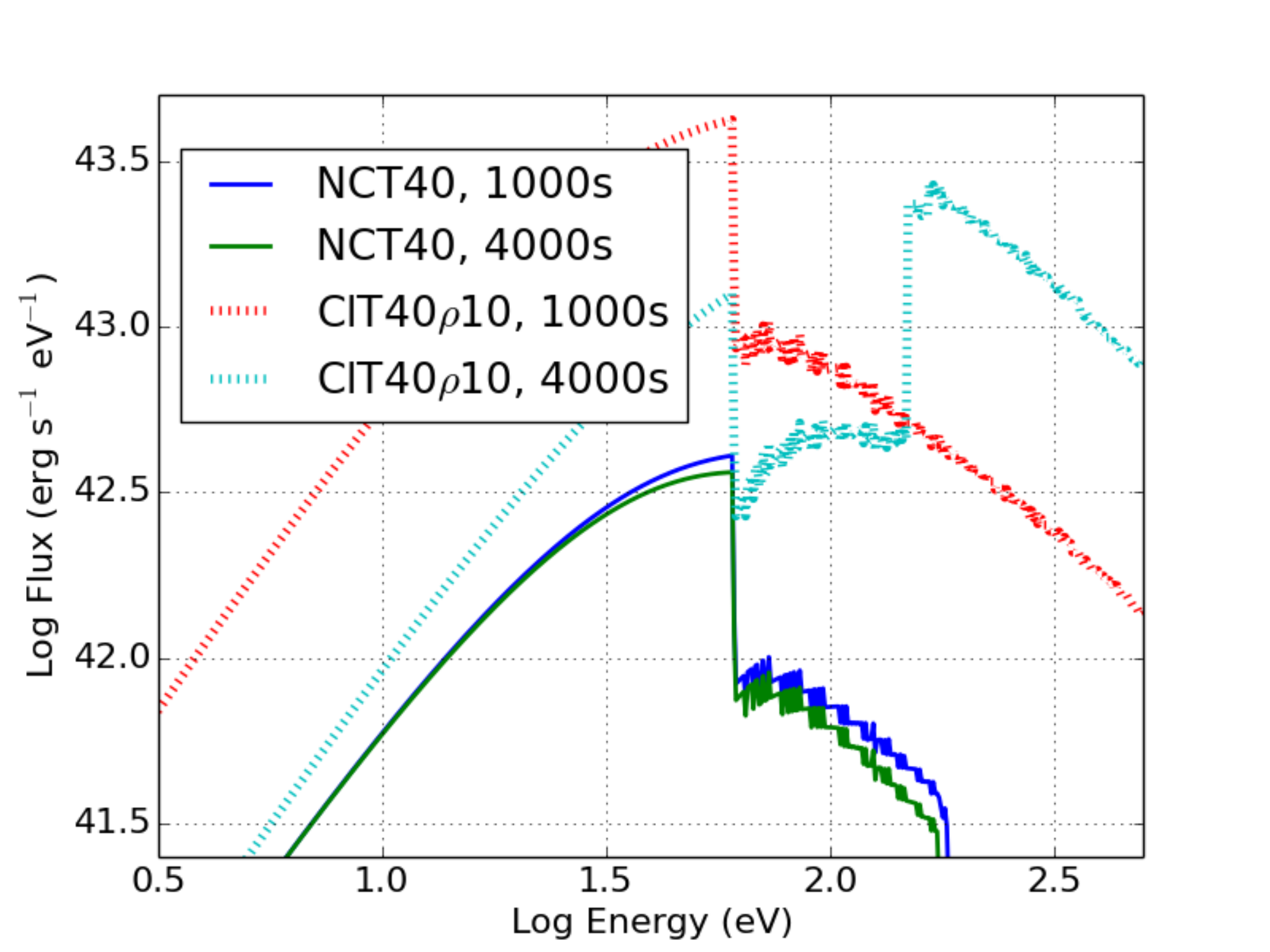}
\caption{Spectra at two different times for a pure wind profile versus our standard clumpy wind profile:  CT40M1.  The shock heating against the clumps produces much hotter material that dramatically increases the high energy emission.  Above 100~eV, a number of line features appear in the spectrum. }
\label{fig:specclump}
\end{figure}

The effects of shocks is even more dramatic in our clumpy wind medium.  Figure~\ref{fig:specclump} shows the spectra comparing a simple $r^{-2}$ wind medium to that of a clumpy wind medium.  The additional shock heating in the clumps leads to considerably higher high-energy emission.  These spectra were calculated by using a single ray trace through our simulations.  When we compared different ray traces, the results were comparable (but with some variability due to variability in the hot spots) and the sum of the ray traces would produce spectra similar to that presented here.  Above 100\,eV, a number of line features appear in our solar abundance wind.

\begin{figure}[ht!]
\plotone{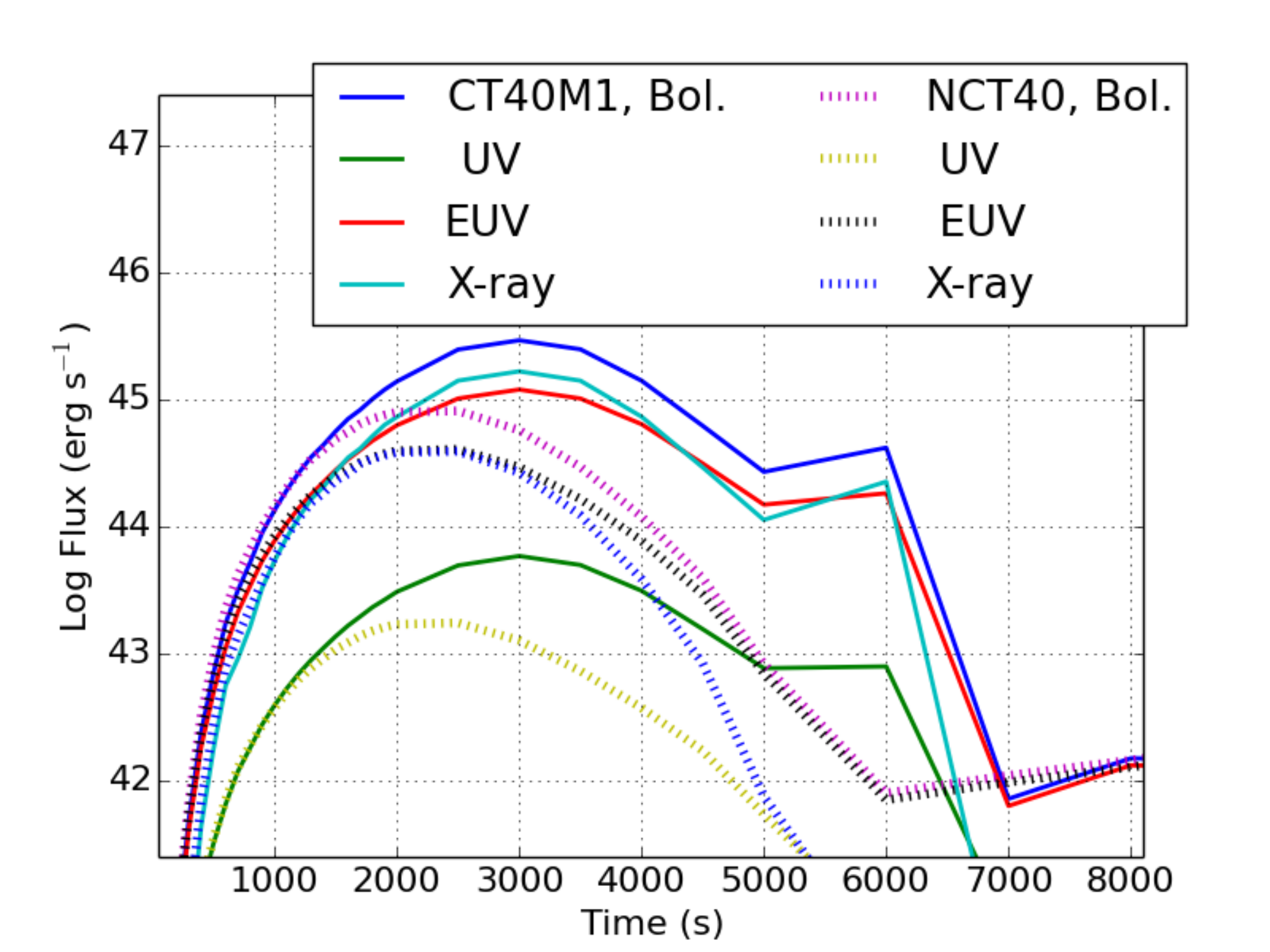}
\caption{Bolometric, UV, extreme UV, and X-ray luminosities as a function of time comparing our $r^{-2}$ wind to our standard clump (CT40M1) simulation.  Although the UV emission is these models are very similar, the Extreme UV emission in our clumpy wind simulation is slightly higher.  The X-ray emission in our clumpy wind simulation is nearly 2 orders of magnitude higher and persists for nearly 4000~s (versus less than 1000~s in the pure wind case).  The noise in the late-time emission occurs because of hot clumps falling in and out of the ray trace.}
\label{fig:lumcnccomp}
\end{figure}

\begin{figure}[ht!]
\plotone{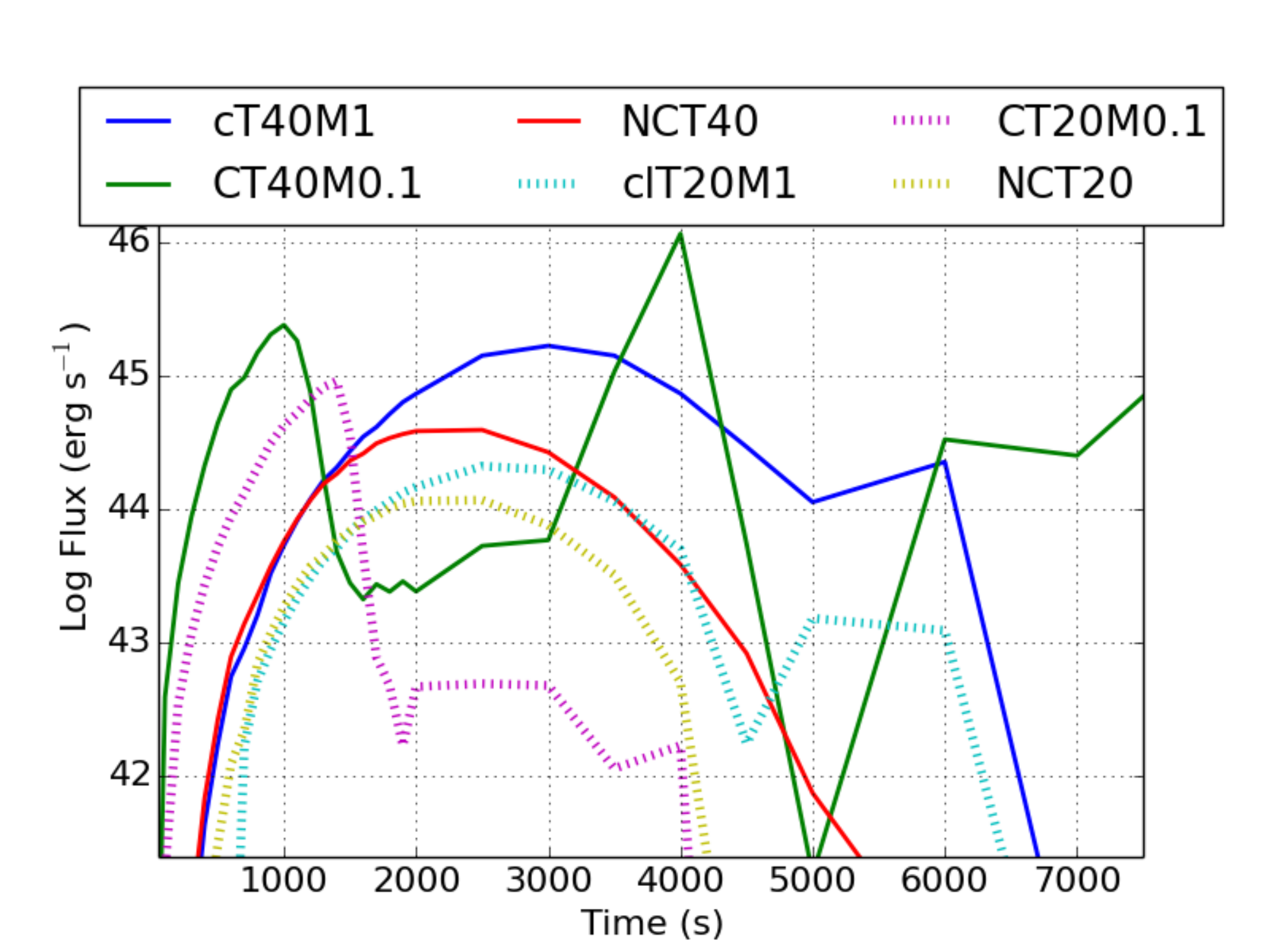}
\caption{X-ray fluxes versus time for our 20 and 40~eV drive temperatures:  CT40M1, CT40M1, NCT40, CT20M1, CT20M1, NCT20.  Variability in the clumps and clump interactions leads to large variability in the X-ray fluxes.}
\label{fig:xray}
\end{figure}

The effect of these hot spots is readily observed in the light-curves with the differences being greatest in the X-rays.  Whereas the light-curve of the simple wind model is strongest in the Extreme UV, it is dominated by the X-rays in the clumpy model.  The X-rays in the clumpy model peak 2 orders of magnitude higher than than the simple $r^{-2}$ wind (Fig.~\ref{fig:xray}).  The strong X-ray emission lasts for nearly 4000~s in the clumpy model versus the 1000~s in the pure wind. As with the shell models,  the peak UV and X-ray luminosities, peak luminosity timescale and width of the light-curves for all of our clumpy models is shown in Table~\ref{tab:results}.  It is clear that the nature of shock breakout will depend sensitively on the nature of the circumstellar medium.  Models that assume a simple $r^{-2}$ wind profile can drastically underestimate the breakout emission.

\begin{figure}[ht!]
\plotone{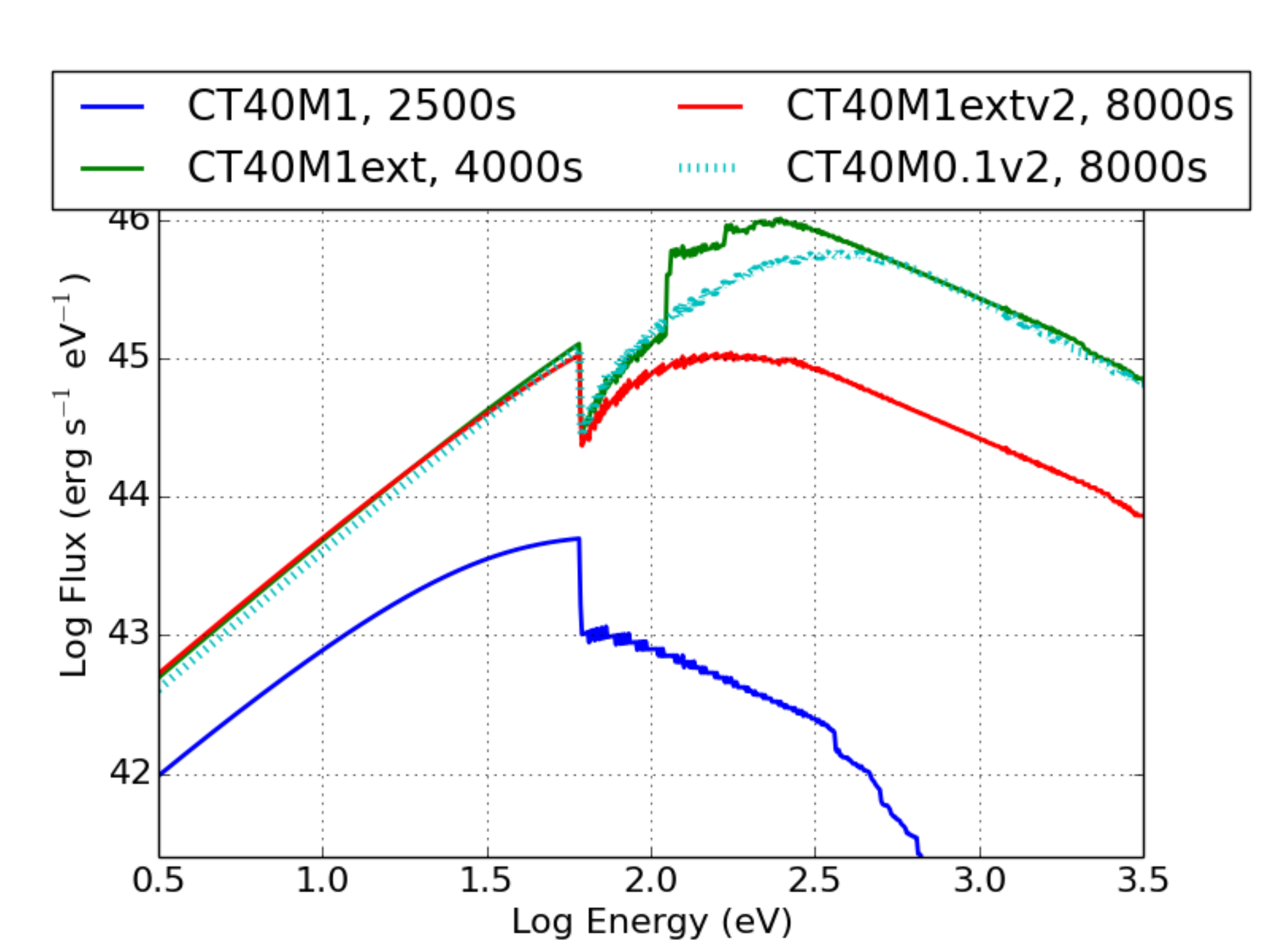}
\caption{Spectra from clumpy models with both increased shock velocities and more extended clump distributions to compare to our standard models.  With higher velocities and more extended clumps, the emission can extend above a keV.}
\label{fig:xrayext}
\end{figure}

With our concentrated clump distributions (out to $5.1\times10^{12}$~cm) and velocities (10,000 ${\rm km \, s^{-1}}$), the spectral fluxes drop off precipitously at 300~eV.  But if we increase the velocity or extend the clumps further into the wind, we produce higher X-ray energies (easily above 1~keV).  The fluxes of some of these extended and high velocity models is shown in Figure~\ref{fig:xrayext}.  The X-ray flux, especially, is very sensitive to the clump distribution and supernova blast wave velocity.

\begin{table*}
  \centering
  \scriptsize
  \begin{tabular}{lcccccccccccc}
  \hline
  \hline
Model & $L_{\rm bol}^{\rm peak}$  & $T_{\rm bol}^{\rm peak}$ & $\Delta T_{\rm bol}$ & $L_{\rm UV}^{\rm peak}$  & $T_{\rm UV}^{\rm peak}$ & $\Delta T_{\rm UV}$ & $L_{\rm EUV}^{\rm peak}$  & $T_{\rm EUV}^{\rm peak}$ & $\Delta T_{\rm EUV}$ & $L_{\rm X-ray}^{\rm peak}$  & $T_{\rm X-ray}^{\rm peak}$ & $\Delta T_{\rm X-ray}$ \\
& (erg s$^{-1}$) & (s) & (s) & (erg s$^{-1}$) & (s) & (s) & (erg s$^{-1}$) & (s) & (s) & (erg s$^{-1}$) & (s) & (s) \\
\hline
NCT40 &  8.15E+44 & 2.50E+03 & 2.20E+03 & 1.75E+43 & 2.50E+03 & 2.70E+03 & 4.06E+44 & 2.50E+03 & 2.70E+03 & 3.92E+44 & 2.50E+03 & 2.20E+03 \\
NCT20 & 2.57E+44 & 2.50E+03 & 2.20E+03 & 7.09E+42 & 2.50E+03 & 2.30E+03 & 1.33E+44 & 2.50E+03 & 2.20E+03 & 1.17E+44 & 2.50E+03 & 2.20E+03 \\ 
NCT10 & 4.28E+44 & 2.50E+03 & 2.60E+03 & 1.17E+43 & 2.50E+03 & 2.60E+03 & 2.06E+44 & 2.50E+03 & 2.60E+03 & 2.10E+44 & 2.50E+03 & 2.50E+03 \\
\hline
CT40M1 & 2.93E+45 & 3.00E+03 & 2.60E+03 & 5.87E+43 & 3.00E+03 & 2.70E+03 & 1.20E+45 & 3.00E+03 & 2.70E+03 & 1.67E+45 & 3.00E+03 & 2.60E+03 \\
CT40M1v2 & 1.54E+46 & 2.50E+03 & 2.70E+03 & 1.04E+44 & 2.50E+03 & 4.50E+03 & 2.98E+45 & 2.50E+03 & 3.50E+03 & 8.65E+45 & 2.50E+03 & 2.00E+03 \\
CT40M1ext & 1.88E+49 & 4.00E+03 & 5.00E+02 & 2.82E+44 & 4.00E+03 & 5.00E+02 & 3.97E+46 & 4.00E+03 & 5.00E+02 & 1.73E+49 & 4.00E+03 & 5.00E+02 \\
CT40M1extv2 & 1.64E+48 & 8.00E+03 & 1.00E+03 & 3.01E+44 & 9.00E+03 & 1.00E+03 & 2.98E+46 &  9.00E+03 & 1.00E+03 & 1.42E+48 & 8.00E+03 & 8.00E+03 \\
CT40M0.1 & 1.18E+46 & 4.00E+03 & 5.00E+02 & 3.61E+43 & 1.00E+03 & 7.00E+02 & 8.48E+44 & 1.00E+03 & 7.00E+02 & 1.15E+46 & 4.00E+03 & 5.00E+02 \\
CT40M0.1v2 & 1.22E+49 & 8.00E+03 & 1.00E+03 & 2.34E+44 & 8.00E+03 & 2.00E+03 & 4.24E+46 & 8.00E+03 & 2.00E+03 & 1.14E+49 & 8.00E+03 & 1.00E+03 \\
CT40M0.1extv2 & 2.04E+48 & 3.00E+03 & 5.00E+02 & 2.45E+44 & 3.00E+03 & 1.00E+02 & 2.33E+46 & 3.00E+03 & 5.00E+02 & 1.87E+48 & 3.00E+03 & 5.00E+02 \\
CT20M1 &  4.45E+44 & 2.50E+03 & 2.40E+03 & 1.26E+43 & 2.50E+03 & 2.50E+03 & 2.21E+44 & 2.50E+03 & 2.50E+03 & 2.10E+44 & 2.50E+03 & 2.40E+03 \\
CT20M0.1 & 1.39E+45 & 1.40E+03 & 6.00E+02 & 3.42E+43 & 1.40E+03 & 7.00E+02 & 4.42E+44 & 1.40E+03 & 7.00E+02 & 9.10E+44 & 1.40E+03 & 6.00E+02 \\

CT10M1 & 9.81E+44 & 3.00E+03 & 2.50E+03 & 3.02E+43 & 3.00E+03 & 2.60E+03 & 4.59E+44 & 3.00E+03 & 2.60E+03 & 4.92E+44 & 3.00E+03 & 2.00E+03\\

\hline
sh0.1r5T40 & 1.12E+46 & 1.90E+03 & 3.00E+03 & 9.02E+43 & 1.90E+03 & 3.20E+03 & 2.17E+45 & 1.90E+03 & 3.00E+03 & 8.96E+45 & 1.90E+03 & 3.00E+03 \\
sh0.1r1T40 & 1.84E+46 & 1.00E+02 & 1.00E+02 & 1.02E+45 & 2.00E+03 & 1.40E+03 & 1.84E+46 & 1.00E+02 & 1.00E+02 & 3.98E+44 & 2.00E+03 & 1.50E+03 \\
sh0.1r2T40 &  2.26E+47 & 6.00E+03 & 1.00E+03 & 7.68E+43 & 2.50E+03 & 2.50E+03 & 4.12E+45 & 6.00E+03 & 5.00E+02 & 2.22E+47 & 6.00E+03 & 1.00E+03 \\
sh1r5T40 & 3.06E+45 & 3.50E+03 & 2.50E+03 & 6.87E+43 & 3.50E+03 & 2.50E+03 & 1.23E+45 & 3.50E+03 & 2.50E+03 & 1.76E+45 & 3.50E+03 & 2.50E+03 \\
sh1r1T40 & 5.73E+45 & 3.50E+03 & 2.00E+03 & 5.15E+44 & 3.50E+03 & 2.00E+03 & 5.06E+45 & 3.50E+03 & 2.00E+03 & 1.64E+44 & 3.50E+03 & 2.00E+03 \\
sh1r2T40 & 3.68E+45 & 5.00E+03 & 4.00E+03 & 7.49E+43 & 5.00E+03 & 4.00E+03 & 1.46E+45 & 5.00E+03 & 4.00E+03 & 2.15E+45 & 5.00E+03 & 4.00E+03 \\
sh1r5T10 & 4.86E+45 & 3.50E+03 & 5.50E+03 & 7.04E+44 & 3.50E+03 & 5.00E+03 & 4.12E+45 & 3.50E+03 & 5.50E+03 & 3.06E+43 & 3.50E+03 & 2.50E+03 \\

\hline
\hline

  \end{tabular}
  \caption{UV, EUV and X-ray bands are, respectively, 3--10~eV, 10--100~eV, and above 100~eV.}
  \label{tab:results}
\end{table*}

\section{Conclusions}

In this paper, we show how inhomogeneities in the circumstellar medium can drastically alter the shock heating in supernova shock breakout and produce a wide range of emission models.  These models give a hint at the wide range of emission spectra and light-curves we can expect as we increase the number of shock breakout observations.  Although this implies that shock breakout observations are ideally suited to probe the immediate surroundings of supernova progenitors, the ability of shock breakout signal to measure the stellar radius is limited by these effects.

The X-ray fluxes from our set of models range from $10^{43}-10^{45} {\rm \, erg \, s^{-1}}$, appearing to match the XRT observations of SN2008D.  Many of our models do not produce the high-energies observed in SN2008D. However, the results depend sensitively on the nature of the clumps and harder X-rays are definitely possible, even with the same properties of the supernova blastwave (velocities and temperatures).  In addition, if the blastwave is even just 2 times faster than our canonical value of 10,000${\rm \, km \, s^{-1}}$, we can produce much higher fluxes at higher photon energies.  This paper touches just the tip of the iceberg of the importance of an inhomogeneous medium on the shock breakout signal.  More detailed studies are in needed to systematically study the full extent of inhomogeneites and comparing the effects of the inhomogeneities to other quantities scientists would like to constrain with shock breakout observations:  stellar radii, mass-loss, and supernova blastwave properties.  But it is clear from this work that it may be possible for clumpy media to explain some of the extreme "shock breakout" observations.

It is also worth mentioning the deficiencies in the simulations.  Our simulations did not model a complete supernova explosion, using instead a boundary source drive to mimic the evolution of the supernova blastwave.  Because of shock interactions within the star, the exact nature of the supernova blastwave is not fully understood and our simple boundary conditions are well within the uncertainties of realistic calculations.  However, leveraging detailed calculations of supernova blastwave calculations and systematic studies of the sensitivity to these boundary conditions to these blastwave conditions is important.  We also defer this study to later studies.

Although state-of-the-art, the models used here made a number of simplifying assumptions.  Although the radiation transport is done using a higher-order-scheme with multigroup opacities, the opacities themselves were calculated assuming a single temperature for the electrons (that dominate collisional excitation) and radiation (driving radiative excitation).  Out of equilibrium effects may well change the atomic level states and the resultant atomic opacities.  This physics must be studied and its effects constrained to produce accurate models.  All of our calculations and results assume the emission is thermal (Kirchoff's law applies).  Non-thermal emission (e.g. synchrotron) could dominate the emission and will likely produce higher-energy emission than our thermal sources.  Finally, we focused on a pencil-beam (Cartesian-grid), 2-dimensional simulation. Spherical grid calculations in 3-dimensions are ultimately necessary to model the shock break-out signal at high precision. 

\bibliography{refs}{}
\bibliographystyle{aasjournal}

\end{document}